\documentclass[a4paper,twocolumn,aps,prx]{revtex4}

\newcommand{\eq}[1]{\begin{equation}#1\end{equation}}
\newcommand{\dd}{\mathrm{d}}
\newcommand{\ee}{\mathrm{e}}

\newcommand{\Tr}{\mathrm{Tr}}
\newcommand{\rp}{\mathrm{Re \,}}
\newcommand{\ip}{\mathrm{Im \,}}

\newcommand{\acos}{\mathrm{acos}}

\newcommand{\atan}{\mathrm{atan}}
\newcommand{\twn}[1]{\langle \mathcal{T}_n(#1)\rangle}
\usepackage{amsmath}
\usepackage{amsfonts}
\usepackage{graphics}
\usepackage{graphicx}
\usepackage{psfrag}
\usepackage{color}
\usepackage{colordvi}
\usepackage{bbm}

\begin{document}

\title{Front dynamics and entanglement in the XXZ chain with a gradient}

\author{Viktor Eisler and Daniel Bauernfeind}
\affiliation{
Institut f\"ur Theoretische Physik, Technische Universit\"at Graz, Petersgasse 16,
A-8010 Graz, Austria
}

\begin{abstract}
We consider the XXZ spin chain with a magnetic field gradient and study the
profiles of the magnetization as well as the entanglement entropy. For a slowly varying field
it is shown that, by means of a local density approximation, the ground-state magnetization
profile can be obtained with standard Bethe ansatz techniques. 
Furthermore, it is argued that the low-energy description of the theory is given by a Luttinger
liquid with slowly varying parameters. This allows us to obtain a very good approximation
of the entanglement profile using a recently introduced technique of conformal field theory
in curved spacetime. Finally, the front dynamics is also studied after the gradient field
has been switched off, following arguments of generalized hydrodynamics for integrable systems.
While for the XX chain the hydrodynamic solution can be found analytically, 
the XXZ case appears to be more complicated and the magnetization
profiles are recovered only around the edge of the front via an approximate numerical solution.

\end{abstract}

\maketitle

\section{Introduction}

Nonequilibrium dynamics of quantum many-body systems continues to
be one of the most actively developing areas of condensed matter physics
\cite{PSSV11,EFG15}. Within this vast field, the studies of integrable systems
have received a particular attention \cite{CEM16}. To a certain degree, this is due
to the powerful analytical tools of integrability, devised during the last century \cite{Baxter,KBI},
which could be further extended to attack a variety of out-of-equilibrium problems.
More importantly, integrable quantum systems turn out to be physically very
interesting, as they show rather peculiar relaxational behaviour \cite{VR16}.
Last but not least, the spectacular development of cold-atom experiments \cite{BDZ08},
making the preparation, control and observation of near-integrable systems feasible
\cite{LGS16}, provided an ultimate boost to the theoretical research activities.

One standard approach to these nonequilibrium studies is a homogeneous global quench,
where the initial state is prepared in the ground state of a certain Hamiltonian
and time evolved with another one, differing in some global parameter, yet both of
them being translational invariant \cite{EF16}.
Indeed, many essential features of anomalous relaxation and thermalization,
resulting from the highly nontrivial families of conservation laws in integrable
systems \cite{IMPZ16}, could be understood in this simple context.
Recently, however, the focus of interest has been shifted towards
the study of inhomogeneous initial states, which might lead to the buildup of
persistent macroscopic currents. An important milestone in this context
has been the introduction of a generalized hydrodynamic picture \cite{BCDNF16,CADY16},
which extends the framework of classical hydrodynamics by taking into
account the entire set of conservation laws in integrable systems.
Starting from an inhomogeneous initial state, the method has been very
successful in describing the time-evolved profiles of various observables (e.g. spin or energy
densities and currents) in a hydrodynamic scaling regime,
for a number of different situations and model systems
\cite{DY17,DSY17,DS17,DYC17,DDKY17,BVKM17a,BVKM17b,PDNCBF17,IDN17b}.

The simplest example of an inhomogeneity in the context of spin
chains is a domain wall, i.e. when two parts of a system
are prepared initially in their (otherwise homogeneous) ground states
with different magnetizations. The resulting front structures have been
widely studied for the XX \cite{ARRS99,HRS04,PK07,AKR08,ER13,AHM14,VSDH15,ADSV16},
the transverse Ising \cite{Karevski02,ZGEN15,EME16,PG17,Kormos17}
and the XXZ chains \cite{GKSS05,AHM14,KMHM14,ZGEN15,BCDNF16}.
In the latter case, a remarkable analytical solution has been found 
very recently \cite{CDLV17}, using the method of generalized hydrodynamics (GHD).
Another commonly studied example of inhomogeneous initial states
is the tensor product of two Gibbs states at different temperatures,
which produces nontrivial energy density and current profiles under time evolution,
see the recent reviews \cite{BD16,VM16} and references therein.

In all of the above mentioned cases the inhomogeneity is sharp
and there is no intrinsic length scale involved in the problem.
Consequently, the time-evolved profiles of magnetization are scaling functions
of $x/t$, i.e. the distance from the location of the initial domain wall divided by time,
reflecting the ballistic nature of the transport. This is not true any more if the
inhomogeneity is localised in an interface region of size $\xi$, with
a smoothly varying magnetization profile. Such an initial state
can be prepared by applying a magnetic field gradient along the chain,
which is then turned off to monitor the subsequent time evolution
\cite{EIP09,LM10,SM13,VIR17}.
In particular, the case of the gradient XX chain with a linearly varying field
can be solved exactly \cite{EIP09}, and  the profiles turn out to be very closely
related to the domain-wall problem.

Here we perform a similar study of the gradient XXZ chain, or equivalently,
interacting spinless fermions in a linear chemical potential.
Although the inhomogeneous potential breaks the integrability of the model, 
we shall show that, for small gradients, the ground-state magnetization profiles
can still be perfectly captured by the Bethe ansatz method combined
with a local density approximation. Moreover, it is demonstrated that the
gradient also gives rise to nontrivial entanglement profiles.
Relying on a recently introduced technique for the study of entanglement
in inhomogeneous free-fermion ground states \cite{DSVC17,DSC17},
we apply a perturbative extension of the method, which is shown
to give good results for a broad regime of interactions in the gradient XXZ chain.
Finally, we also monitor the time evolution of the profiles after switching off
the gradient. Invoking the GHD method, an explicit analytical solution 
of the hydrodynamical density profile is obtained in the XX case.
For the XXZ chain the method is applied with certain simplifications,
that are shown to give a good numerical approximation in the edge
region of the front.

The paper is structured as follows. In Sec. \ref{sec:model} we introduce
the model and its basic features, continued by an analysis of the
ground-state magnetization profiles in Sec. \ref{sec:mag}. 
The entanglement profiles are studied in Sec. \ref{sec:ent},
with a brief introduction into the curved-space CFT technique 
for free fermions and its perturbative extension to the XXZ case.
The front dynamics after switching off the gradient is investigated in Sec. \ref{sec:front}.
We conclude the paper in Sec. \ref{sec:disc} by a discussion of our
results against recent developments. Some details of the
calculations are enclosed in two Appendices.

\section{XXZ chain with a gradient\label{sec:model}}

The Hamiltonian of the XXZ chain is given by
\eq{
H_{XXZ} = \sum_{j=-L/2+1}^{L/2-1} (S^x_{j}S^x_{j+1} +S^y_{j}S^y_{j+1} +
\Delta S^z_{j}S^z_{j+1} )
\label{hxxz}}
where the spin-$1/2$ operators $S^{\alpha}_j=\sigma^{\alpha}_j/2$
on site $j$ are given through Pauli matrices with $\alpha=x,y,z$,
and $\Delta$ is the interaction strength. The XXZ chain with a gradient
is obtained by adding a linearly varying magnetic field
\eq{
H_{\mathrm{gr}} = H_{XXZ} - \delta h \sum_{j=-L/2+1}^{L/2} (j-1/2)S^z_j \, ,
\label{hgr}}
with the slope set by $\delta h$.
The gradient field is shifted such that it has a zero between sites $j=0$
and $j=1$, and thus the Hamiltonian is invariant under the spin-inversion
$S^z_j \to - S^z_{1-j}$. Note that, due to the special geometry of the problem,
it is natural to consider open boundary conditions. Performing a Jordan-Wigner
transformation, the gradient XXZ chain is equivalent to a fermionic hopping
chain with nearest-neighbour interactions and a linear chemical potential.
%
%

Our first goal is to characterize the inhomogeneous ground state of $H_{\mathrm{gr}}$
via the magnetization and entanglement profiles. The inverse of the gradient
defines a length scale $\xi=\delta h^{-1}$. For $|j|\gg \xi$, the magnetization
is saturated at the values $\langle S^z_j \rangle = \pm 1/2$ on the far right- and
left-hand sides of the chain, whereas the two parts are connected by
a nontrivial interface region.
Although $H_{XXZ}$ is solvable by Bethe Ansatz, the gradient field breaks the
integrability of the Hamiltonian. Nevertheless, we shall try to capture the
properties of the interface by using standard Bethe Ansatz methods
in combination with a local density approximation (LDA). The essential argument
behind LDA is that, if the field varies slowly enough, $\delta h \ll 1$, then the
properties of the inhomogeneous ground state will also change smoothly.
In particular, at site $j$ the spins experience a local field strength
$h(j) = \delta h (j-1/2)$, and the system will locally occupy the ground state
corresponding to a \emph{homogeneous} field $h=h(j)$. In other words,
if the length scale $\xi$ associated with the gradient is much larger then the
lattice spacing, LDA should give a good approximation of the profiles.

\section{Magnetization profiles\label{sec:mag}}

%
\begin{figure*}[t]
\center
\includegraphics[width=\columnwidth]{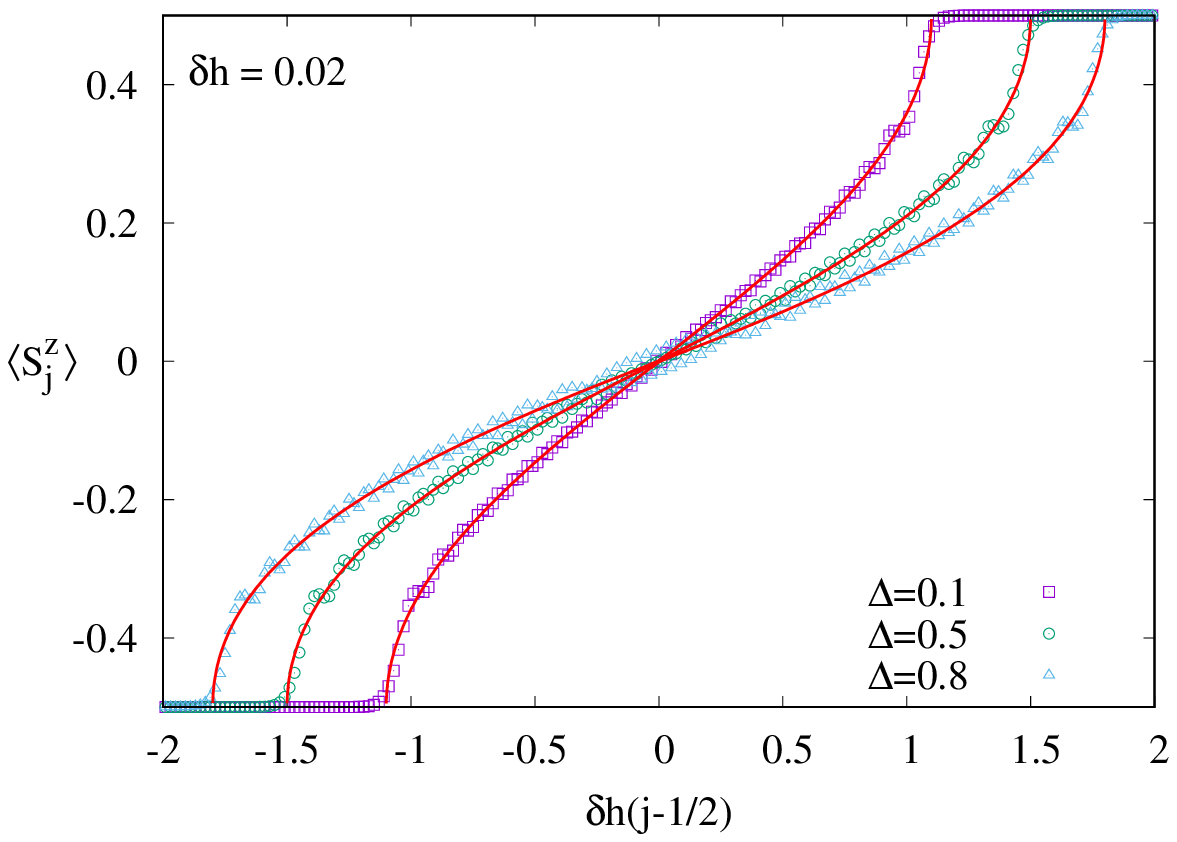}
\includegraphics[width=\columnwidth]{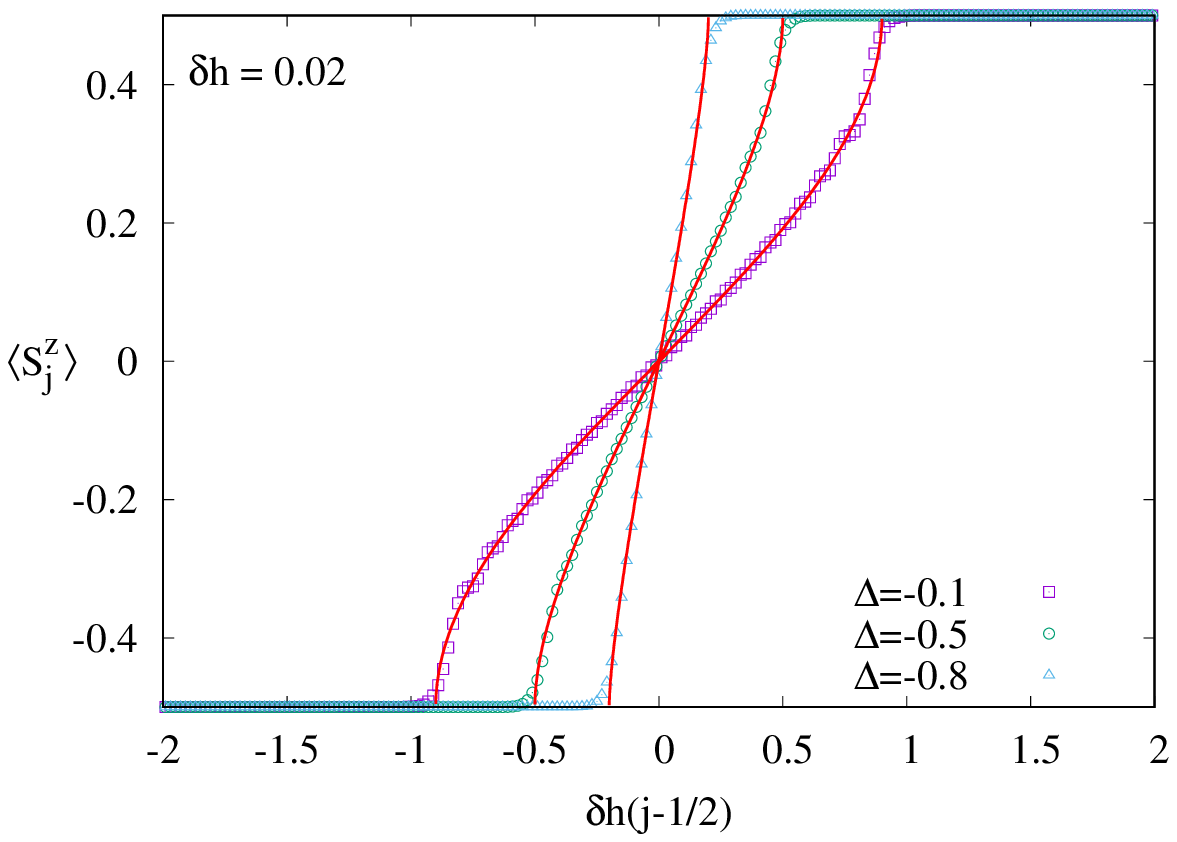}
\caption{Magnetization profiles obtained from DMRG (symbols) vs. TBA solution (red lines).
Calculations for a chain of $L=200$ with magnetic field gradient $\delta h=0.02$, plotted
against $\delta h(j-1/2)$.}
\label{fig:sz}
\end{figure*}
%

Obtaining the magnetization profiles for the gradient problem thus boils down
to solving for the ground state of the XXZ chain in a nonzero field $h$.
We will focus on the planar regime $|\Delta|<1$ and use the standard
parametrization $\Delta = \cos \gamma$. 
Working directly in the limit $L \to \infty$, this is a well-known problem
in terms of the thermodynamic Bethe Ansatz (TBA) \cite{MT,FF}. Indeed, the
eigenstates of the XXZ chain are parametrized by rapidities $\lambda$,
corresponding to magnon excitations created upon the fully polarized state.
In particular, the ground state has a simple Fermi sea character, with real
rapidities fully occupying an interval $\lambda \in \left[-\Lambda,\Lambda \right]$.
While for finite $L$ the rapidities satisfy some appropriate quantization
conditions, in the thermodynamic limit the roots become continuous, and their
density $\rho(\lambda)$ follows from the TBA equation \cite{MT,FF}
\eq{
\rho(\lambda) +
\frac{1}{2\pi} \int_{-\Lambda}^{\Lambda} 
\mathcal{K}(\lambda-\mu) \rho(\mu) \dd \mu=
\frac{ k'(\lambda)}{2\pi} ,
\label{rho}}
where the derivative of the bare momentum
$k'(\lambda) \equiv \theta'_1(\lambda)$ and the integral kernel
$\mathcal{K}(\lambda) \equiv \theta'_2(\lambda)$ are given through
\eq{
\theta'_n(\lambda) = \frac{\sin n\gamma}
{\cosh \lambda -\cos n\gamma} \,.
\label{tpn}}

Since the magnons are created on the ferromagnetic state, the
local magnetization can be obtained as
\eq{
\langle S^z \rangle = \frac{1}{2} - \int_{-\Lambda}^{\Lambda}
\rho (\lambda) \dd \lambda \, .
\label{sz}}
Hence, for a fixed Fermi rapidity $\Lambda$, the magnetization
is a simple function of the root density, which in turn follows from
the integral equation \eqref{rho}. However, to extract the profiles
one needs the magnetization as a function of the magnetic field $h$.
To obtain the relation $h(\Lambda)$, one first defines the so-called
dressed charge function $Z(\lambda)$, which is the solution of yet
another integral equation
\eq{
Z(\lambda) + \frac{1}{2\pi} \int_{-\Lambda}^{\Lambda} 
\mathcal{K}(\lambda-\mu) Z(\mu) \dd \mu = 1 \, .
\label{Z}}
In terms of the root density and the dressed charge, the magnetic
field is then given by \cite{MT,FF}
\eq{
h(\Lambda) = \frac{2\pi\rho(\Lambda)}{Z(\Lambda)}
\sin \gamma \, .
\label{hlam}}

Unfortunately, the analytical solutions of $\rho(\lambda)$ and $Z(\lambda)$
are available only in the limiting cases $\Lambda = \infty$,
corresponding to $\langle S^z \rangle = h =0$,
and $\Lambda=0$ which yields $\langle S^z \rangle =1/2$.
In the latter case the value of the critical magnetic field $h_c$
follows from the trivial solutions $2\pi\rho(0)=\theta'_1(0)$ and $Z(0)=1$ as 
\eq{
h_c = 
\frac{\sin^2 \gamma}{1-\cos{\gamma}} = 1+\Delta \, .
}
If the magnetic field exceeds the value $h_c$, the ground state
remains fully polarized. Consequently, within the validity of the LDA framework 
for the gradient chain, the interface connecting the ferromagnetic regions
is located between $-h_c < h(j) < h_c$, and its half-width is given by
$\xi(1+\Delta)$.

The complete profile can be obtained via the numerical solutions of the TBA
equations \eqref{rho} and \eqref{Z}, by gradually increasing $\Lambda$
and calculating the matching pairs of $\langle S^z \rangle$ and $h$ from
\eqref{sz} and \eqref{hlam}. The LDA profile can then be compared to the
numerical value of $\langle S^z_j \rangle$ in a finite size chain,
obtained via density matrix renormalization group (DMRG) simulations
\cite{Schollwoeck11,itensor}. The discarded weight was set to $10^{-12}$
without any restriction on the maximal bond dimension,
and we required the ground-state energy to converge within $\Delta E < 10^{-7}$
for three consecutive DMRG sweeps. The results of these calculations
are shown on Fig. \ref{fig:sz}, for a size $L=200$ and field gradient $\delta h = 0.02$.
One can see that, up to some oscillations, the LDA method gives a perfect
description of the profile, both in the antiferromagnetic ($\Delta>0$, left)
as well as in the ferromagnetic ($\Delta<0$, right) regimes.
In the fermionic language, the repulsive or attractive nature of the interactions
yields a stretched or squeezed profile as compared to the non-interacting
case where the half-width of the interface is given by $\xi$ \cite{EIP09}.
Interestingly, the match between the LDA profiles and the data remains
very good even for $\Delta=-0.8$, where the interface is rather sharp.
On the other hand, the oscillations around LDA are more pronounced for
large $\Delta > 0$.

Finally, one observes in Fig. \ref{fig:sz} that the profiles show a nonanalytic
behaviour at the edge of the interface. This can be captured by using
the approximate solutions of the integral equations \eqref{rho} and \eqref{Z}
in the limit $\Lambda \ll 1$, which yields for the magnetic field \cite{FF}
\eq{
h \approx h_c - \frac{1}{\tan^2 \gamma/2} \frac{\Lambda^2}{2} \, .
\label{hhc}}
Furthermore, the integral in Eq. \eqref{sz} can also be approximated as
$\Lambda \theta'_1(0)/\pi$ such that the magnetization becomes
\eq{
\langle S^z \rangle \approx \frac{1}{2} -  \frac{1}{\pi}\sqrt{2(h_c-h)}\, .
\label{szhc}}
Substituting $h \to h(j) = (j-1/2)/\xi$, one can see that the profile $\langle S^z_j \rangle$
shows a square-root singularity at the edge, which is independent of the
interaction parameter $\Delta$.

\section{Entanglement profiles\label{sec:ent}}

We continue by analyzing the entanglement profiles in the ground state.
We are primarily interested in the behaviour of the von Neumann entropy
\eq{
S = - \Tr \, \rho_A  \ln \rho_A, \qquad
\rho_A = \Tr_B |\psi\rangle\langle\psi| \, ,
\label{ent}}
where $\rho_A $ is the reduced density matrix for a bipartition of the
ground state $|\psi\rangle$. We consider only the case of
intervals $A=\left[-L/2+1,r\right]$ and $B=\left[r+1,L/2\right]$, where
$r$ measures the distance of the cut from the centre of the chain.
In particular, $r=0$ corresponds to the half-chain case and the
entropies for $\pm r$ must be exactly equal, due to
the symmetry of the ground state under the combined action of reflection and
spin-flip transformations.

Although the entanglement entropy is obtained straightforwardly
from DMRG simulations, it is much harder to develop an analytical
method, based on some LDA argument, that could provide accurate
predictions for the entropy profile. A recent breakthrough has been achieved
for free-fermion systems by realizing that, in a proper continuum limit, the
inhomogeneous problem can be mapped onto a CFT in a curved space \cite{DSVC17}.
Our goal is to use this approach as a starting point for the more complicated
interacting problem of the gradient XXZ chain,
we shall thus first review the main ideas and arguments of the method.

\subsection{Dirac field theory in curved space \label{sec:cft}}

Let us consider a free-fermion Hamiltonian in 1D continuous space
with a slowly varying potential term $V(x)$
\eq{
H = \int_{-\infty}^{\infty} \dd x \, c^{\dag}(x)
\left[ -\partial^2_x -\mu +V(x)\right] c(x) \, ,
\label{hff}}
where $c^{\dag}(x)$ and $c(x)$ are the creation/annihilation operators.
With $V(x) \equiv 0$, the constant chemical potential $\mu$ sets the ground
state to be a Fermi sea between the Fermi momenta $\pm k_F$, where
from the dispersion $\varepsilon(k)=k^2$ one has
$k_F=\sqrt{\mu}$. Within the realms of LDA, the main effect of a slowly
varying potential $V(x)$ is to induce a space-dependent Fermi momentum
$k_F(x)=\sqrt{\mu-V(x)}$. In particular, the fermionic density $\rho(x)=k_F(x)/\pi$
is nonvanishing only in a domain $V(x) < \mu$.

Introducing the time-evolved fermionic operator
$c(x,y)=\ee^{yH}c(x)\ee^{-yH}$ with imaginary time $y=it$, a more
precise description of the local spacetime behaviour of the theory is
encoded in the fermionic propagator
\eq{
\langle c^{\dag}(x+\delta x,y+ \delta y)c(x,y)\rangle \simeq
\frac{i}{2\pi} \left( \frac{\ee^{-i\delta \varphi}}{\delta z} -
\frac{\ee^{i\delta \bar\varphi}}{\delta \bar z}\right),
\label{ffprop}}
where we introduced
\eq{
\delta z = \delta x + iv_F(x) \delta y \, , \quad
\delta \varphi = k_F(x) \delta x + i v_F(x) \delta y \, .
}
Here the Fermi velocity is defined as
$v_F(x)=\varepsilon'(k_F(x))$ and its space dependence is
entirely due to the variation of the Fermi momentum $k_F(x)$. 
Note that the form \eqref{ffprop} of the propagator is valid only
in the local neighbourhood of spacetime point $(x,y)$, as the
derivation involves a linearization of the dispersion $\varepsilon(k)$
around $k_F(x)$.

The main idea of Ref. \cite{DSVC17} is to view the two terms of
Eq. \eqref{ffprop} as the chiral propagators of a 2D massless
Dirac field. In a flat Euclidean spacetime, this is defined by the action
\eq{
\mathcal{S} = \frac{1}{\pi} \int \dd z \dd \bar z
\left [\psi_R^{\dag} \partial_{\bar z} \psi_R +
\psi_L^{\dag} \partial_z \psi_L \right],
\label{dirac}}
where $z=x+iv_Fy$ and the right- and left-moving
chiral propagators are given by
\eq{
\langle \psi_R^{\dag}(z)\psi_R(0)\rangle = \frac{1}{z}, \quad
\langle \psi_L^{\dag}(z)\psi_L(0)\rangle = \frac{1}{\bar z}.
\label{chprop}}
Comparing to \eqref{ffprop}, one indeed finds the expected
\emph{local} behaviour, up to some phase factors which
can be incorporated by a chiral gauge transformation.
The crucial question is, however, whether there exists a
proper choice of complex coordinates $w(x,y)$, such that
it provides a \emph{globally} valid Dirac action in a curved
space?

It turns out that for this to be the case, the underlying Riemannian
metric should satisfy \cite{DSVC17}
\eq{
\dd s^2 = \ee^{2\sigma} \dd w \, \dd \bar w \, ,
\label{ds2}}
so that the chiral propagator behaves locally as
\eq{
\langle \psi_R^{\dag}(w+\delta w)\psi_R(w)\rangle=
\frac{1}{\ee^\sigma \delta w} \, .
}
Hence, to be able to derive Eq. \eqref{ffprop} from a
Dirac theory in curved space, one should have
\eq{
\ee^{\sigma} \delta w(x,y) = \delta x + iv_F(x) \delta y \, .
}
It is easy to see that this condition is satisfied by the choice
\eq{
w(x,y) = \int^{x} \frac{\dd x'}{v_F(x')} + iy \, , \qquad
\ee^{\sigma} = v_F(x) \, .
\label{wxy}}
Note that this map involves only the transformation of
spatial coordinates, whereas the imaginary time is left untouched.
Moreover, it depends on a single parameter,
namely the local Fermi velocity $v_F(x)$.

%
\begin{figure*}[htb]
\center
\includegraphics[width=\columnwidth]{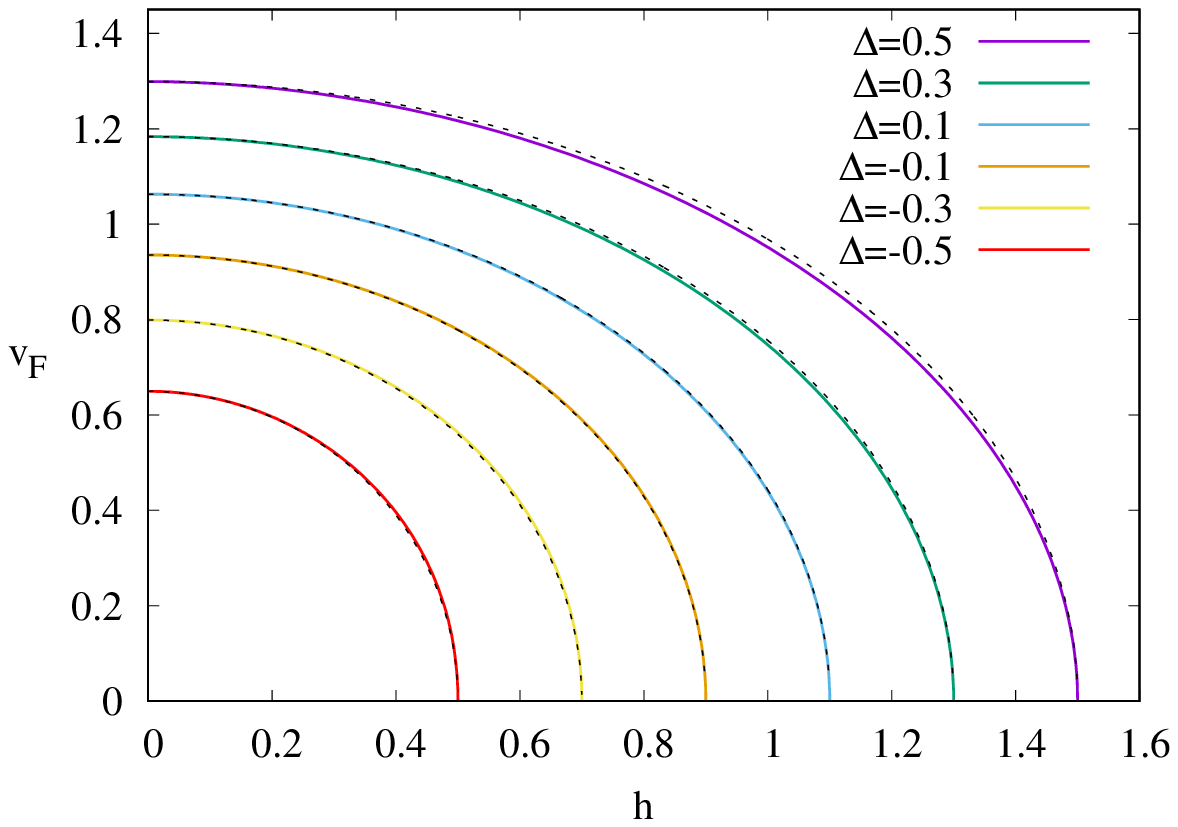}
\includegraphics[width=\columnwidth]{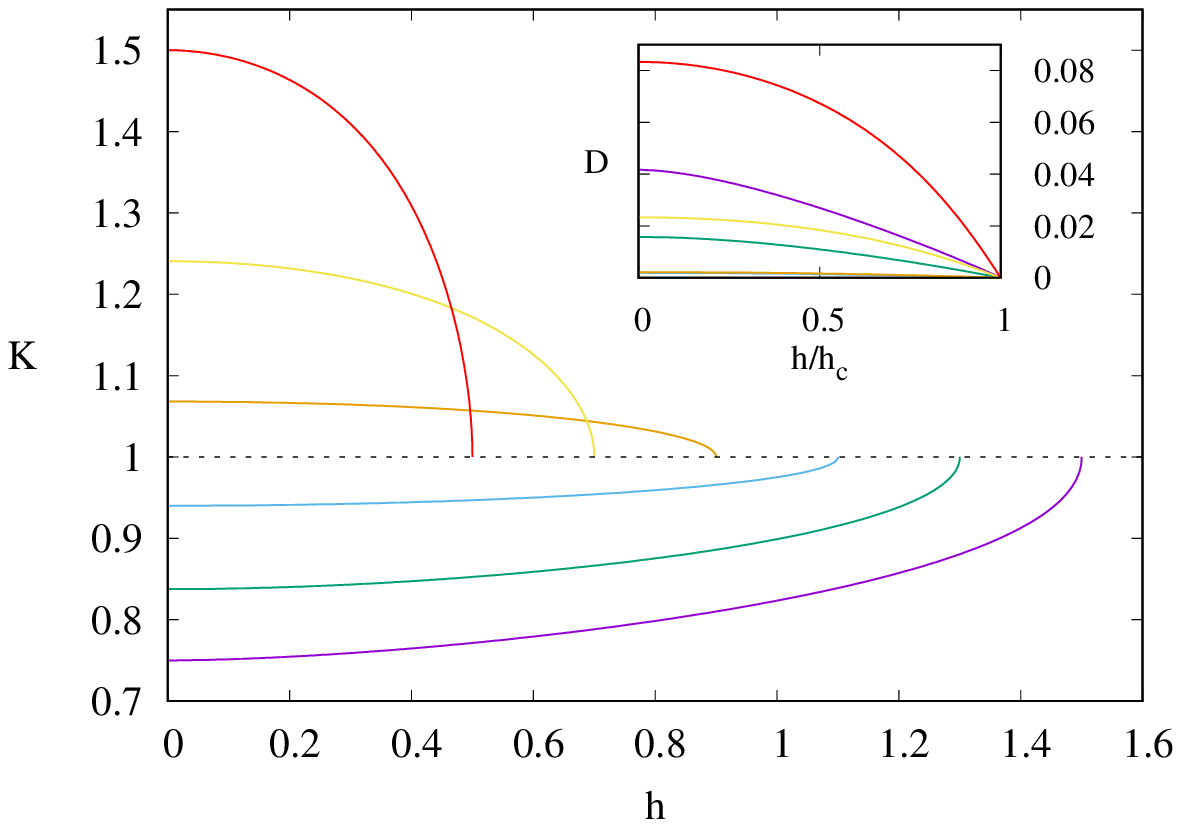}
\caption{Fermi velocity (left) and Luttinger parameter (right) as obtained from 
TBA, see Appendix \ref{app:vfk}. The black dashed lines in the left part show
the approximation Eq. \eqref{vfh}. The inset on the right shows the scaling
dimension $D$ in Eq. \eqref{D}.}
\label{fig:vfk}
\end{figure*}
%

\subsection{Inhomogeneous Luttinger liquids}

The case of the XXZ chain, or equivalently interacting fermions,
turns out to be more complicated.  Indeed, the low-energy behaviour
of the model is described, after standard bosonization procedure \cite{TG},
by a Luttinger liquid
\eq{
H_{LL} =\int \dd x  \frac{v_F}{2} \left[ K (\partial_x \theta)^2
+ K^{-1} (\partial_x \phi)^2 \right],
\label{hll}}
which is an effective field theory involving the bosonic field
$\phi$ and its dual $\theta$, satisfying the standard bosonic
commutation relation
$\left[ \phi(x), \partial_{x'}\theta(x')\right] = i \delta(x-x')$.
Clearly, apart from the Fermi velocity $v_F$, we have now the
extra Luttinger parameter $K$ that enters the field theory description.
It should be noted that, while bosonization yields both of these
parameters only within perturbation theory, the exact values of
$v_F$ and $K$ can be fixed from the Bethe ansatz solution.
This procedure is well-known for the XXZ chain with a
homogeneous magnetic field  \cite{TG,Sirker12,FF},
with the results shown in Fig. \ref{fig:vfk}.
The method is summarized in Appendix \ref{app:vfk}.

Clearly, when considering the gradient XXZ chain,
the variation of the parameters $v_F(h)$ and $K(h)$
w.r.t. the magnetic field translates into a spatial variation,
with the identification $h\to x = j/\xi$ (from here on, the $-1/2$
shift in the $j$ coordinate will be suppressed for brevity of notation).
Hence, the field-theory
description is still given by a Luttinger liquid as in Eq. \eqref{hll},
however with parameters $v_F(x)$ and $K(x)$ that vary slowly
with position, on a length scale given by $\xi$.
The situation is thus more complicated than the one considered
recently in Ref. \cite{DSC17}, where the case of variable Fermi velocity
$v_F(x)$ but uniform Luttinger parameter $K$ was dealt with.
This might occur in inhomogeneous
problems such as the XXZ chain with a varying coupling $J(x)$.
Indeed, in this case the parameter $K$ can essentially be scaled
out from the Luttinger liquid Hamiltonian \eqref{hll} by a canonical
transformation $\phi \to \sqrt{K} \phi$ and $\theta \to \theta/\sqrt{K}$.
On the level of CFT, the rescaling of the bosonic fields amounts
to changing the compactification radius of the theory.
Nevertheless, the variation of $v_F(x)$ can still be incorporated
in a curved metric just the same way as was done in \eqref{wxy}
for the Dirac action \cite{DSC17}.

The inhomogeneous Luttinger liquid problem, with both parameters
$v_F(x)$ and $K(x)$ being nonuniform, is hard in general.
Instead of trying to find an exact solution, we shall rather be interested
in uncovering a perturbative regime, where the Dirac field theory in
curved space yields a good approximation. Having a look at
the behaviour of $K(h)$ in the right of Fig. \ref{fig:vfk}, one observes
that the variation is rather strong already for relatively small
interactions $\Delta$. At first sight, this would opt against a perturbative
approach. However, to better understand the role played by the
Luttinger parameter, one should have a look at the fermionic propagator.
From standard results of bosonization \cite{TG}, the propagator
of the right-moving fermion has the form
\eq{
\langle \psi_R^{\dag}(z+\delta z)\psi_R(z)\rangle \propto
\frac{1}{\delta z} \frac{1}{|\delta z|^D} \, ,
\label{xxzprop}}
where the proportionality sign indicates that we have
omitted the phase factor and normalization constant, cf. Eq. \eqref{ffprop}.
Similarly, the left-moving propagator is obtained by substituting $\delta z \to \delta \bar z$.

The local deviation from a Dirac theory is thus encoded in the
scaling dimension
\eq{
D(x) = \frac{K(x)+K^{-1}(x)}{2}-1 \, .
\label{D}}
Obviously, setting $K(x) \equiv 1$ we obtain $D(x) \equiv 0$ and recover the local Dirac
propagators in \eqref{chprop}. Moreover, the result \eqref{xxzprop} suggests
that, whenever the scaling dimension varies slowly and satisfies $D(x) \ll 1$
in the full spatial domain, one should expect the curved-space Dirac theory to yield
a good approximation to the inhomogeneous Luttinger liquid problem.
Remarkably, as shown in the inset on the right of Fig. \ref{fig:vfk}, 
this condition is indeed satisfied for a broad range of $\Delta$.
In particular, one has $D \to 0$ as $h \to h_c$ for arbitrary values of $\Delta$ \cite{CHP98}.
In general, $D(x)$ increases towards the bulk and reaches a
maximum deviation of approx. $4\%$ for $\Delta=0.5$, resp.
$8\%$ for $\Delta=-0.5$. Clearly, deviations increase further as $|\Delta|\to1$
and one leaves the perturbative regime.

%
%

\subsection{Entropy and curved-space CFT}

%
\begin{figure*}[htb]
\center
\includegraphics[width=\columnwidth]{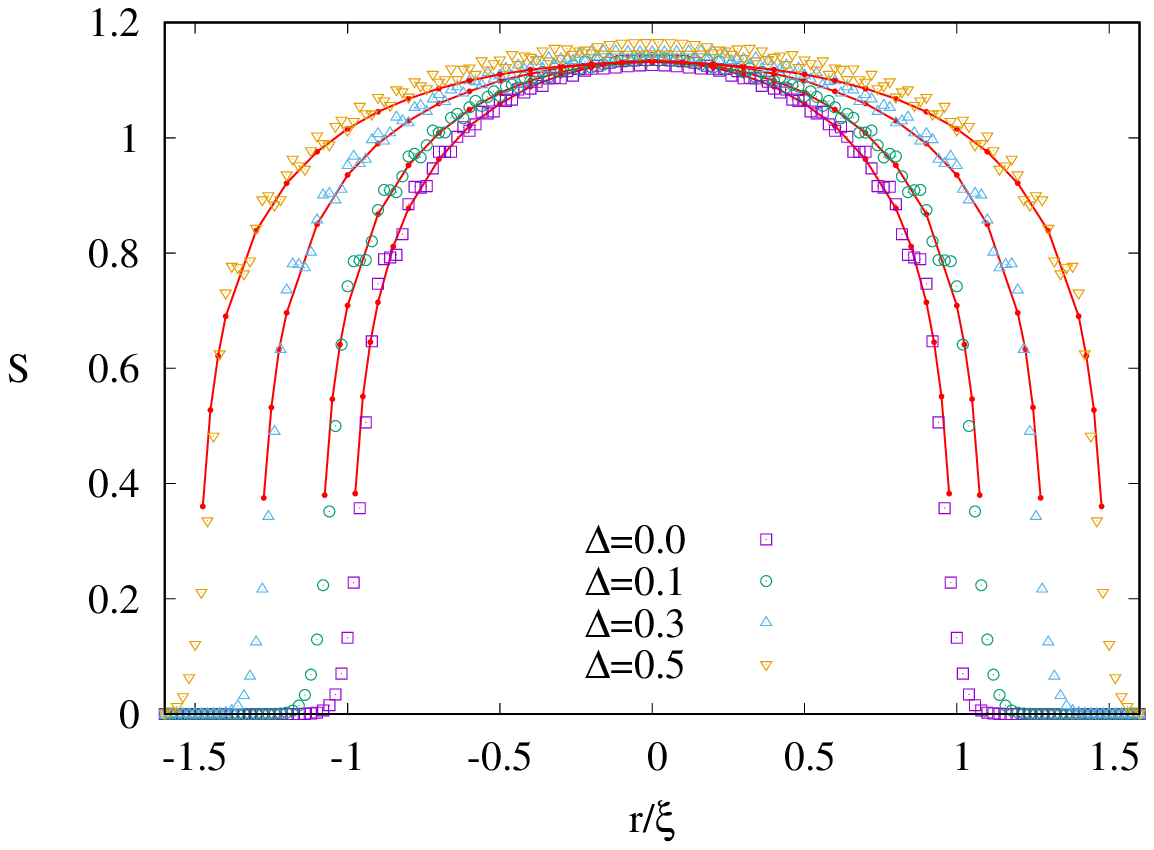}
\includegraphics[width=\columnwidth]{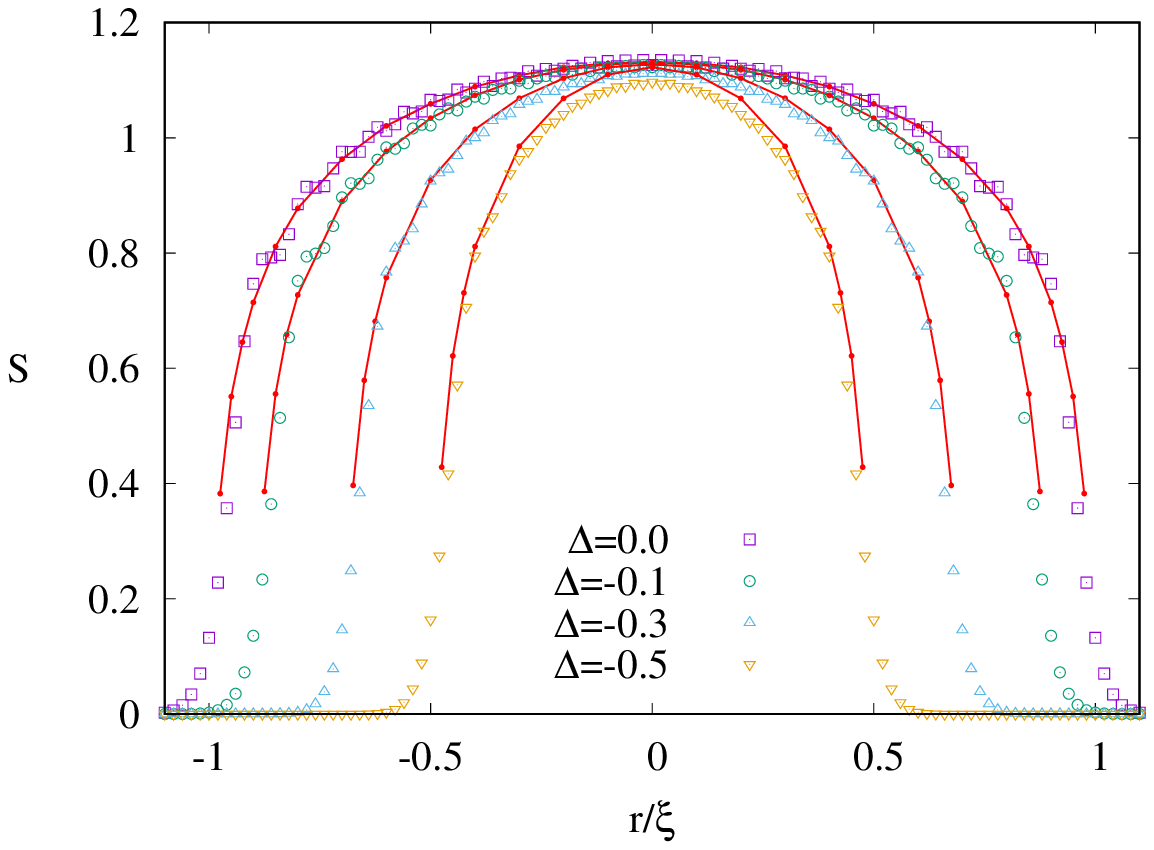}
\caption{Entanglement entropy obtained from DMRG (symbols) vs. curved-space CFT approach (red dotted lines).
Calculations for a chain of $L=200$ with magnetic gradient $\delta h=0.02$.}
\label{fig:ent}
\end{figure*}
%

The validity of the above arguments will be tested by comparing the
predictions of the curved-space CFT approach for the entanglement
entropy to the DMRG results. Once the curved metric is fixed by
\eqref{wxy}, the entropy can be extracted by calculating expectation
values of twist-field operators \cite{CC09} and applying conformal
transformations to simplify the CFT geometry \cite{DSVC17}.
The steps of this procedure are collected in Appendix \ref{app:cft}.
It should be noted that, while there is no exact analytical expression
for $v_F(h)$, we found that, for intermediate values of $\Delta$,
the behaviour of the Fermi velocity as a function of the magnetic field
is very well approximated by the ansatz
\eq{
v_F(h) = v_0 \sqrt{1-\left(\frac{h}{h_c}\right)^2}, \qquad
v_0 = \frac{\pi}{2} \frac{\sin \gamma}{\gamma},
\label{vfh}}
where $v_0$ is the Fermi-velocity at zero magnetization.
The ansatz, shown by the dashed lines in the left of Fig. \ref{fig:vfk},
is the simplest possible generalization of the noninteracting result by
matching the analytically known values $v_F(0)=v_0$ and $v_F(h_c)=0$.
Using this approximation in \eqref{wxy} and following the recipe in
Appendix \ref{app:cft}, one finds that the entropy of a segment
$\left[-L/2+1,r\right]$ is given by
\eq{
S = \frac{1}{6} \ln \mathcal{L}(r) + C(r,\Delta) \, ,
\label{s}}
where the conformal distance reads
\eq{
\mathcal{L}(r) = 
h_c \xi \left[1-\left(\frac{r}{h_c \xi} \right)^2\right] .
\label{l}}

Unfortunately, the curved-space CFT calculation fixes the entropy only up to
some non-universal (i.e. depending on the underlying lattice model)
term $C(r,\Delta)$, which still contains a dependence on the position $r$ of the cut.
In fact, this is completely analogous to standard CFT calculations of the
ground-state entropy in the XXZ chain with a homogeneous field $h$.
There the entropy of a segment of length $\ell$ has the form
\eq{
S_{\mathrm{hom}} = \frac{1}{6} \ln \tilde{\mathcal{L}}(\ell) + \tilde{C}(h,\Delta) \, ,
\label{shom}}
with the universal term given by the chord length
\eq{
\tilde{\mathcal{L}}(\ell) = \frac{L}{\pi} \sin \left(\frac{\ell\pi}{L}\right) .
\label{lgs}}
From analogy of the expressions \eqref{s} and \eqref{shom},
and in spirit of the LDA argument, one infers that the non-universal
terms are related as
\eq{
C(r,\Delta) = \tilde C(r/\xi,\Delta) \, .
\label{crd}}

In general, the non-universal term must be extracted from the lattice model
and an analytical expression is known only in the free-fermion case \cite{JK04,FC11}
\eq{
\tilde C(h,0) = \frac{1}{6}\ln \sqrt{1-h^2}+ c_0 \, ,
\label{cxx}} 
with a numerical value of the constant $c_0 \approx 0.4785$.
This yields the entropy of the gradient XX chain as
\eq{
S = \frac{1}{6} \ln \left[\xi (1-(r/\xi)^2)^{3/2}\right] + c_0 \, ,
\label{entxx}}
in perfect accordance with the results found in \cite{EP14}.
For $\Delta \ne 0$, the term $\tilde C(h,\Delta)$ must be extracted by fitting the
ground-state entropy with the ansatz \eqref{shom}, with the results shown in
Appendix \ref{app:cft}.

Finally, having obtained both the universal and non-universal
terms, one can check the results agains the DMRG data.
This is shown in Fig. \ref{fig:ent}, plotted against the variable
$r/\xi$, with the red lines showing the interpolated curved-space
CFT ansatz \eqref{s}. The dots on the lines indicate the parameters
$h=r/\xi$, where the actual values of $C(r,\Delta)$ were determined
using the relation \eqref{crd} and data fits for the homogeneous chain.
In general, the ansatz seems to give a very good quantitative description
of the entropy, although some deviations are visible for larger $\Delta$
values. In particular, one observes that the match between \eqref{s}
and the DMRG data is best around the edge of the profile, whereas
deviations increase as one moves towards the bulk. This completely
parallels the behaviour of the scaling dimension $D(x)$
(see inset of Fig. \ref{fig:vfk}) and gives a further confirmation of
our perturbative approach. Note that we have also checked the
result using the exact form of $v_F(h)$ instead of the approximation
\eqref{vfh}. In this case the conformal distance is given by \eqref{l2}
in Appendix \ref{app:cft} and has to be evaluated numerically.
The corresponding deviations induced in \eqref{s} are
of the order of the fitting error in $C(r,\Delta)$ and thus the
approximate form \eqref{l} of the conformal distance is perfectly justified.

\section{Front dynamics\label{sec:front}}

We now proceed to investigate the dynamics that follows after switching off
the gradient magnetic field. The time-evolved state of the system
is then given by
\eq{
|\psi(t) \rangle = \ee^{-itH_{XXZ}} |\psi\rangle \, .
\label{psit}}
As for the static case, we shall study the time-evolved profiles of
magnetization and entanglement.
The problem has a single characteristic length scale $\xi$,
it is thus natural to work with the rescaled coordinate $x=j/\xi$ and time
$\tau = t/\xi$. Since the time-evolution of the state \eqref{psit}
is governed by the integrable XXZ Hamiltonian, we expect the
behaviour of local observables to be given via a generalized hydrodynamic
(GHD) picture introduced recently \cite{BCDNF16,CADY16}.
Within the GHD framework the effective hydrodynamic state of the system
is described by space- and time-dependent occupation functions of rapidities,
that are subject to certain advection equations.
Before formulating the problem in general, it is very instructive to start with
the noninteracting XX case, where an explicit solution of the
GHD equation can be found and compared to the exact lattice solution \cite{EIP09}.

\subsection{Exact solution for the XX chain}

The XX Hamiltonian in fermion language is given by
\eq{
H_{XX} = -\frac 1 2 \sum_{j}(c^{\dag}_j c_{j+1} + c^{\dag}_{j+1} c_{j}) \, ,
\label{hxx}}
and can be diagonalized by a Fourier transform.
Hence, due to the free-fermion nature of the problem,
one can simply use the occupation function $n(x,\tau;k)$ in momentum-space,
instead of working with rapidities. The GHD equation \cite{BCDNF16,CADY16}
for the occupation function then reads 
\eq{
\partial_\tau n(x,\tau;k) + v(k)
\partial_x n(x,\tau;k) = 0 \, ,
\label{advxx}}
where the single-particle velocity is $v(k)=\sin k$.

We assume that the hydrodynamic state of the system is
described, throughout the time-evolution, by an occupation function
\eq{
n(x,\tau;k) = 
\begin{cases}
1 & k \in \left[k_-,k_{+}\right] \\
0 & k \notin \left[k_-,k_{+}\right]
\end{cases},
\label{nkxx}}
which is a local Fermi sea between two Fermi points $k_\pm(x,\tau)$ for any time $\tau$.
Since \eqref{nkxx} depends on $x$ and $\tau$ only via the location of the Fermi sea,
the advection equation \eqref{advxx} yields the equation of motion for the
Fermi points
\eq{
\partial_\tau k_{\pm} + \sin k_{\pm} \partial_x k_{\pm} = 0 \, .
}

One is thus left with two independent advection equations which can be
solved by the method of characteristics. This gives $k_{\pm}$ implicitly,
as the solution of the equation
\eq{
k_{\pm} = F_\pm(x-\tau\sin k_{\pm}) \, ,
\label{kpm}}
where the function $F_{\pm}$ describes the initial conditions
\eq{
F_{\pm}(x) = k_{\pm}(x,0)=\pm\acos \left(x\right), \qquad
|x| \le 1 \, .
\label{kpm0}}
Note that \eqref{kpm0} is simply the LDA result for the gradient
XX chain, where we considered $\delta h < 0$ in Eq. $\eqref{hgr}$
such that the fermionic density
$\langle c^{\dag}_jc_j \rangle=\langle S^z_j \rangle + 1/2$
is zero (one) on the far right (left), and thus the current flows from left to right.
Inverting Eq. \eqref{kpm}, one should look for the solution of 
\eq{
\cos k_{\pm} = x - \tau \sin k_{\pm} \, ,
}
which is a simple quadratic equation with roots
\eq{
\cos k_{\pm} = \frac{x \mp \tau \sqrt{1+\tau^2-x^2}}{1+\tau^2}.
}
Note that real solutions exist only for $|x| \le x_{max}$ and thus
$x_{max}=\sqrt{1+\tau^2}$ gives the location of the front edge.
Using trigonometric identities, the solution can be rewritten in
the simple form
\eq{
k_{\pm}(x,\tau) = 
\pm \acos \frac{x}{\sqrt{1+\tau^2}} + \atan \, \tau \, .
\label{kpmxt}}
Comparing to the initial data \eqref{kpm0}, one sees that,
on top of an overall $x$-independent drift term, the solution
simply gets rescaled by $x_{max}$.

With the solution \eqref{kpmxt} at hand, it is easy to evaluate the expectation
values of the particle density $\rho$ and current $\mathcal J$ in the
hydrodynamic state \eqref{nkxx}. A simple calculation gives 
\eq{
\begin{split}
&\rho = \frac{1}{\pi}\frac{k_{+}-k_{-}}{2}, \\
&\mathcal J = \frac{1}{\pi} \sin \frac{k_{+} - k_{-}}{2} \sin \frac{k_{+} + k_{-}}{2}.
\end{split}
\label{rhoj}}
Substituting the solution \eqref{kpmxt} for the Fermi points, one arrives at
\eq{
\begin{split}
&\rho(x,\tau) = \frac{1}{\pi}\acos \frac{x}{\sqrt{1+\tau^2}}, \\
&\mathcal J(x,\tau) = \frac{1}{\pi} \sqrt{1-\frac{x^2}{1+\tau^2}} \sqrt{\frac{\tau^2}{1+\tau^2}}.
\end{split}
\label{rhojxt}}
Note that the drift term in \eqref{kpmxt} does not enter the expression of the density,
which thus has the same functional form as at $t=0$, up to a rescaling of
the distance by $x_{max}$.
It is easy to check that the particle density
and current in \eqref{rhojxt} satisfy the usual continuity equation
\eq{
\partial_\tau \rho(x,\tau) + \partial_x \mathcal J(x,\tau) = 0 \, .
}
%

The hydrodynamic result \eqref{rhojxt} can now be compared to the 
exact lattice solution of the front dynamics. The lattice density and current
are given via the fermionic correlators $C_{j,l}(t)=\langle c^\dag_j(t) c_l(t) \rangle$ as
\eq{
\rho(j,t) = C_{j,j}(t) \, , \qquad
\mathcal J(j,t) = \ip C_{j,j+1}(t) \, .
\label{rhojjt}}
The explicit form of the correlation matrix was obtained in Ref. \cite{EIP09} and reads
\eq{
C_{j,l}(t) = C^{0}_{j,l}(\sqrt{\xi^2+t^2}) \ee^{i\varphi(l-j)}\, ,
\label{ct}}
where $\varphi = \atan (t/\xi)$ and
\eq{
C^{0}_{j,l}(\xi) = \frac{\xi}{2(j-l)} 
\left[J_{j-1}(\xi) J_{l}(\xi) - J_{j}(\xi) J_{l-1}(\xi) \right]
\label{c0t}}
is the ground-state correlation matrix with Bessel functions $J_j(\xi)$.
Thus, up to a phase factor, the correlators have the exact same form
as in the ground state, with an effective length scale $\sqrt{\xi^2+t^2}$.
In fact, this is the simplest manifestation of a more general phenomenon of
an emergent eigenstate solution, discussed in Ref. \cite{VIR17}.

Furthermore, the solution can also be related to the case, where the initial
state is given by a sharp domain wall (i.e. the initial density is a step function)
by setting $\xi = 0$ in \eqref{ct}. This yields $\varphi=\pi/2$ and
the initial correlations \eqref{c0t} have to be evaluated with an argument $t$.
For the domain-wall case the density and current are well-known scaling functions
of the variable $v=j/t$ and read \cite{ARRS99}
\eq{
\rho_{dw}(v) = \frac{1}{\pi}\acos \, v \, , \quad
\mathcal J_{dw}(v) = \frac{1}{\pi} \sqrt{1-v^2} \, .
}
Then, in the general case $\xi \ne 0$, the scaling form of $\rho$ simply follows by
substituting $t \to \sqrt{\xi^2+t^2}$, since the density is unaffected by the
phase $\varphi$. For the current $\mathcal J$ one has to multiply the result by an
additional factor of $\sin \varphi$, as is clear from \eqref{rhojjt} and \eqref{ct}.
These manipulations lead immediately to the hydrodynamic result \eqref{rhojxt},
which is thus exact in the appropriate scaling limit.

\subsection{Numerical results for $\Delta \ne 0$}

We start the discussion of the interacting case by summarizing
the GHD method for a general integrable system. 
These models possess an extensive set of local (or quasi-local) conserved charges \cite{IMPZ16},
each of them yielding an appropriate continuity equation
that must be respected by the dynamics. Furthermore, within
the TBA description, each of the charge and respective current
densities are fixed by occupation functions $n_l(\lambda)$
where $\lambda$ is the rapidity parameter and $l$ indexes
the families of quasi-particles in the integrable model at hand \cite{IDNWCEP15,IQDNB16}.
In fact, as it was shown in the pioneering works \cite{BCDNF16,CADY16},
these continuity equations can be reformulated as
advective flow equations for the occupation functions
\eq{
\partial_\tau n_l(x,\tau;\lambda) + v_l(\{n\};\lambda)
\partial_x n_l(x,\tau;\lambda) = 0 \, .
\label{adv}}

Although formally very similar to the free-fermion case Eq. \eqref{advxx},
the essential difficulty of the above advection equation is due to the
term $v_l(\{n\};\lambda)$, which is the dressed quasi-particle velocity.
Indeed, as the index $\{n\}$ suggests, it depends in general on the
full hydrodynamic state of the system, therefore coupling the complete
set of advection equations \eqref{adv}.
The precise form of the dressed velocity is fixed within the TBA formalism
for the corresponding model via the dressing equation, see Refs.
\cite{BCDNF16,CADY16} for details.

In the present case, the ground-state results of Sec. \ref{sec:mag} and a
naive analogy to the XX case in \eqref{nkxx} would
suggest that the hydrodynamical description of the dynamics could be
specified by a single occupation function of the form
\eq{
n(x,\tau;\lambda) =
\begin{cases}
1 & \lambda \in \left[\Lambda_-,\Lambda_{+}\right] \\
0 & \lambda \notin \left[\Lambda_-,\Lambda_{+}\right]
\end{cases},
\label{nl}}
where the entire spacetime-dependence is via the Fermi rapidities
$\Lambda_\pm(x,\tau)$. There is, however, an essential difference
between the momentum and rapidity parametrization, which is already
transparent in the $\Delta=0$ case. Namely, the real rapidities yield
only the low-momentum particles $k \le \pi/2$, whereas those
with $k>\pi/2$ correspond to complex rapidities along the $i\pi$ axis \cite{MT}.
In TBA language, the latter correspond to a different family of
quasi-particles with negative parity, and thus the modes are not smoothly
connected to each other. In fact, this is also apparent for the
ground-state at $\tau=0$, where for $x\to 0$ one has $\Lambda_\pm \to \pm \infty$
and the sea of rapidities in \eqref{nl} is already completely filled.

While in the XX case these complications can be avoided by working
in momentum space, for $\Delta \ne 0$ one is bound to run into
difficulties when considering the solution of the advection equations \eqref{adv}.
In general, the quasi-particle structure can be constructed for some particular
values of $\gamma$ (i.e. rational multiples of $\pi$) and is given in terms of
string-solutions for the rapidities \cite{MT}. In analogy to the XX case, the real rapidities yield
only quasi-particles with bare momentum restricted to the interval
$k \in \left[-\gamma,\gamma \right]$, and the particle content at $\tau=0$
and $x<0$ (i.e. in the negative magnetization sector) must be obtained
by looking at the nontrivial action of spin-flip on the quasi-particle structure \cite{DLCDN17,IDN17a}.
However, since the resulting occupation functions $n_l(x,\tau;\lambda)$
are all coupled together via the dressed velocities, even the numerical
solution of \eqref{adv} becomes rather complicated.

Instead of trying to find a full numerical solution to the advection equations,
we shall try to simplify the problem by dealing with only one particle type
of real rapidities ($l=1$), with occupation function as in Eq. \eqref{nl}.
Obviously, such an ansatz will immediately fail around $x \to 0^+$ for any $\tau >0$,
as the quasi-particles from the left penetrate the right half of the system.
Nevertheless, it is reasonable to expect that \eqref{nl} gives a good approximation
of the hydrodynamics around the front edge, at least for small times $\tau$,
where all the other quasi-particle occupations could be neglected.
The equation of motion for the Fermi rapidities then reads
\eq{
\partial_\tau \Lambda_{\pm}  + v^{\pm}_{\left[\Lambda_{-},\Lambda_{+}\right]}
\partial_x \Lambda_{\pm} = 0 \, ,
\label{advlpm}}
where the dressed quasi-particle velocity
\eq{
v^\pm_{\left[\Lambda_{-},\Lambda_{+}\right]} = 
\frac{(\epsilon')^{dr}(\Lambda_\pm)}
{(k')^{dr} (\Lambda_\pm)}
}
is defined via the dressing equation
\eq{
f^{dr}(\lambda) + \frac{1}{2\pi} \int_{\Lambda_-}^{\Lambda_+} 
\mathcal{K}(\lambda-\mu) f^{dr}(\mu) \dd \mu = f(\lambda) \, .
\label{dr}}
Note that the bare momentum derivative $k'(\lambda)=\theta'_1(\lambda)$
is given by Eq. \eqref{tpn}, whereas the bare energy $\epsilon(\lambda)$ is given by
\eqref{epsl} of Appendix \ref{app:vfk}.

\begin{figure}[thb]
\center
\includegraphics[width=\columnwidth]{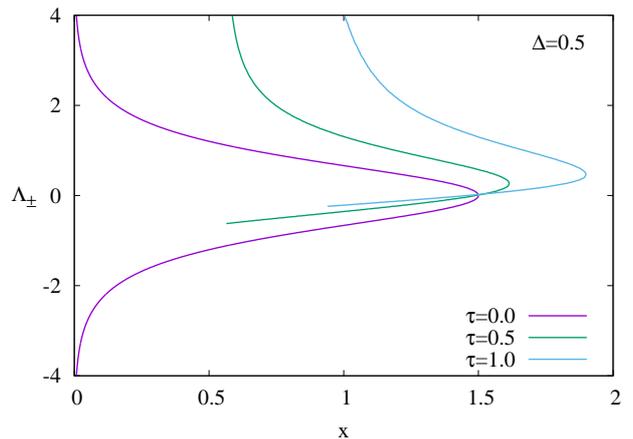}
\caption{Solutions of the advection equation \eqref{advlpm} by method of characteristics \eqref{lpm},
for various times $\tau$ and $\Delta=0.5$.}
\label{fig:lpm}
\end{figure}

In a restricted spatial domain, an approximate solution for $\Lambda_\pm$
can be found again via the method of characteristics, by solving
%
\eq{
\Lambda_{\pm} = F_\pm(x-\tau v^\pm_{\left[\Lambda_{-},\Lambda_{+}\right]}) \, ,
\label{lpm}}
with the initial condition $F_{\pm}(x) = \Lambda_{\pm}(x,0)$.
Inverting this function, we get the initial position $x(\Lambda)=x(-\Lambda)=F^{-1}_{+}(\Lambda)$
as a function of the Fermi rapidity $\Lambda\ge0$. With the identification $x \to h$, this
is nothing else but the function in Eq. \eqref{hlam}, we are thus looking for the
solutions of
\eq{
\frac{2\pi\rho(\Lambda_\pm)}{Z(\Lambda_\pm)}
\sin \gamma 
= x-\tau v^\pm_{\left[\Lambda_{-},\Lambda_{+}\right]} \, .
\label{xlpm}}
Since the left hand side of \eqref{xlpm} is strictly positive,
two distinct solutions exist only in an interval $x_{min}<x<x_{max}$
for any fixed $\tau >0$. In particular, the front position $x_{max}(\tau)$
denotes the coordinate where the solutions $\Lambda_-=\Lambda_+=\Lambda_{*}$
of \eqref{xlpm} coalesce, whereas $x_{min}(\tau)$ corresponds to $\Lambda_+ \to \infty$.
In general, the solutions $\Lambda_\pm$ can be found numerically by an iterative
procedure and are shown on Fig. \ref{fig:lpm}, with an upward drift of the Fermi
rapidities for increasing times.

%
%

%
\begin{figure*}[t]
\center
\includegraphics[width=.66\columnwidth]{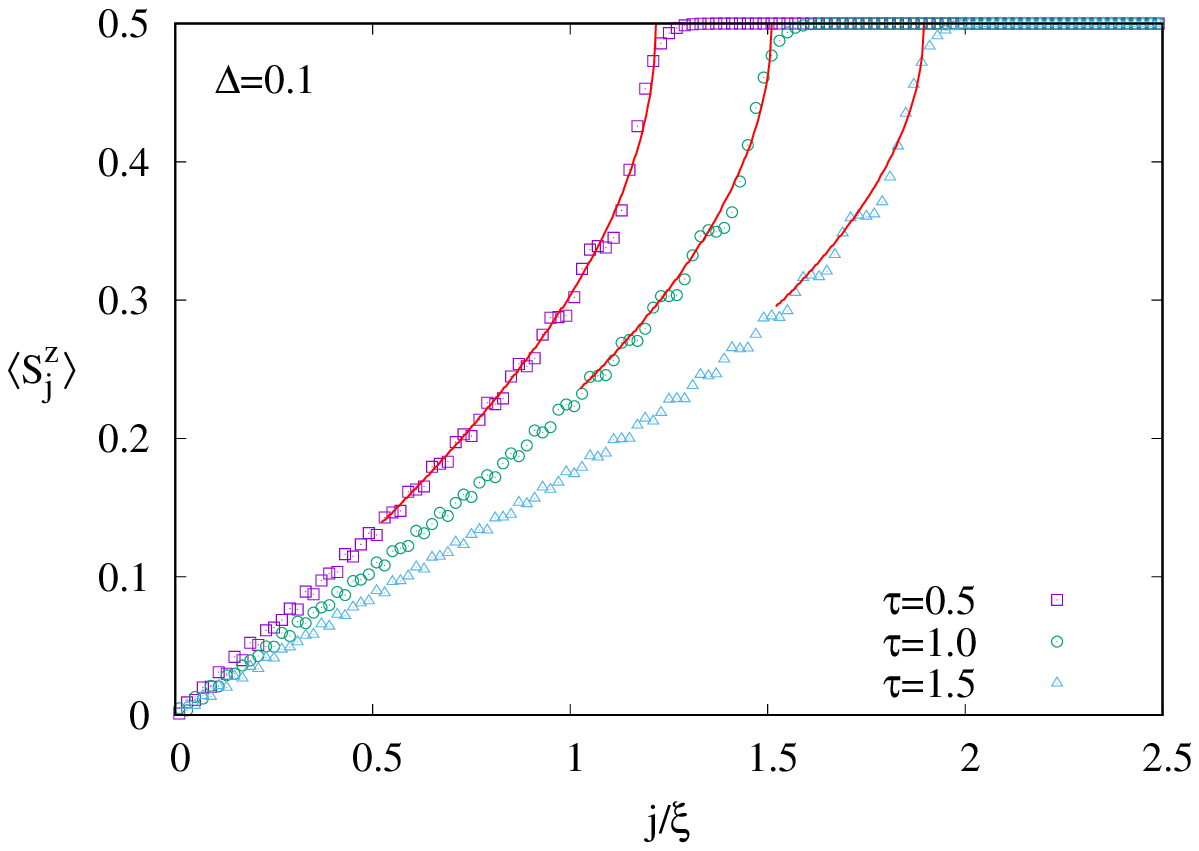}
\includegraphics[width=.66\columnwidth]{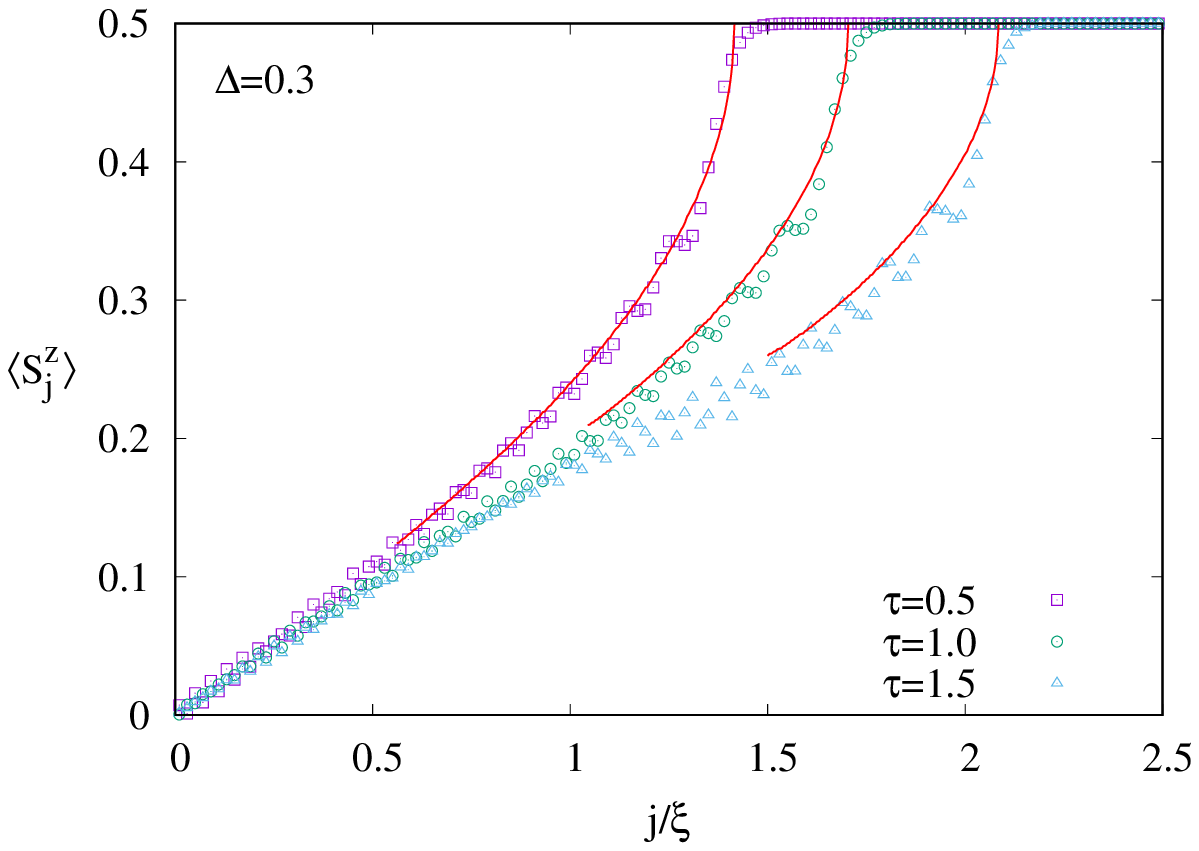}
\includegraphics[width=.66\columnwidth]{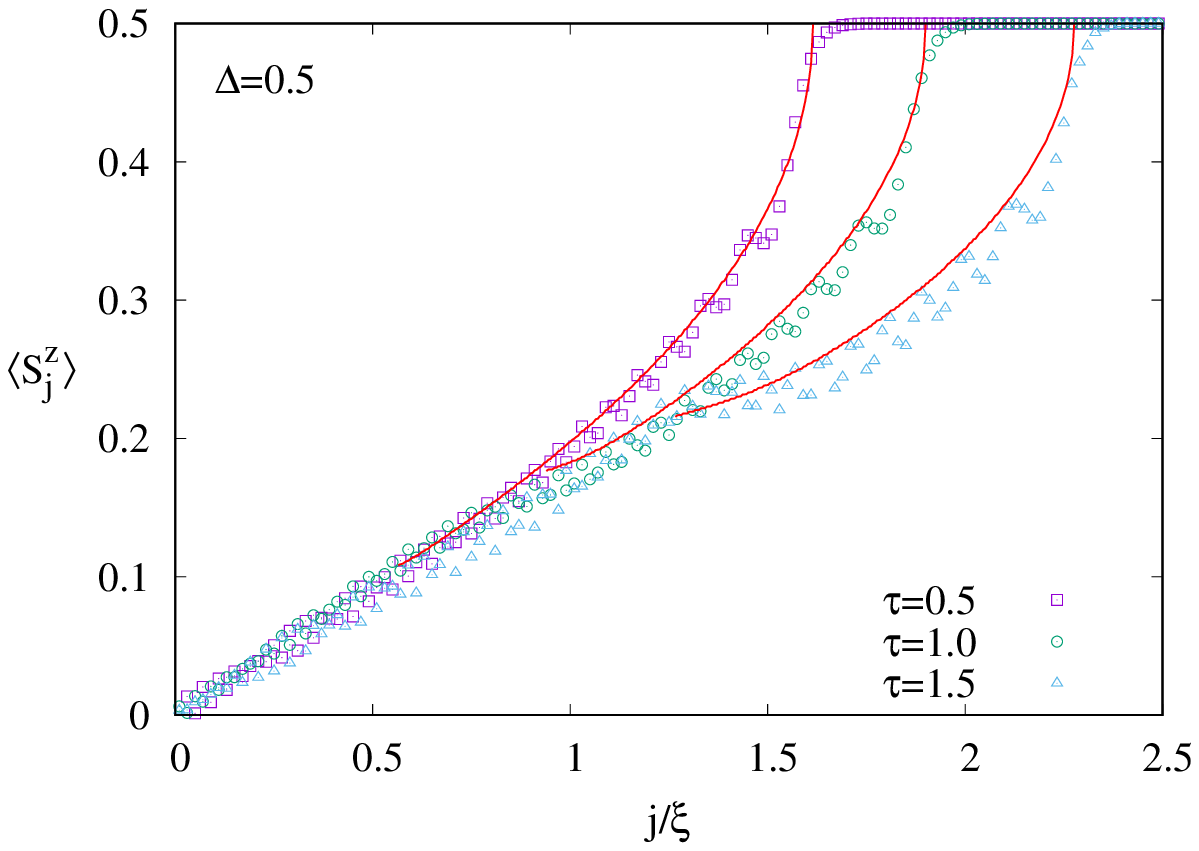}
\includegraphics[width=.66\columnwidth]{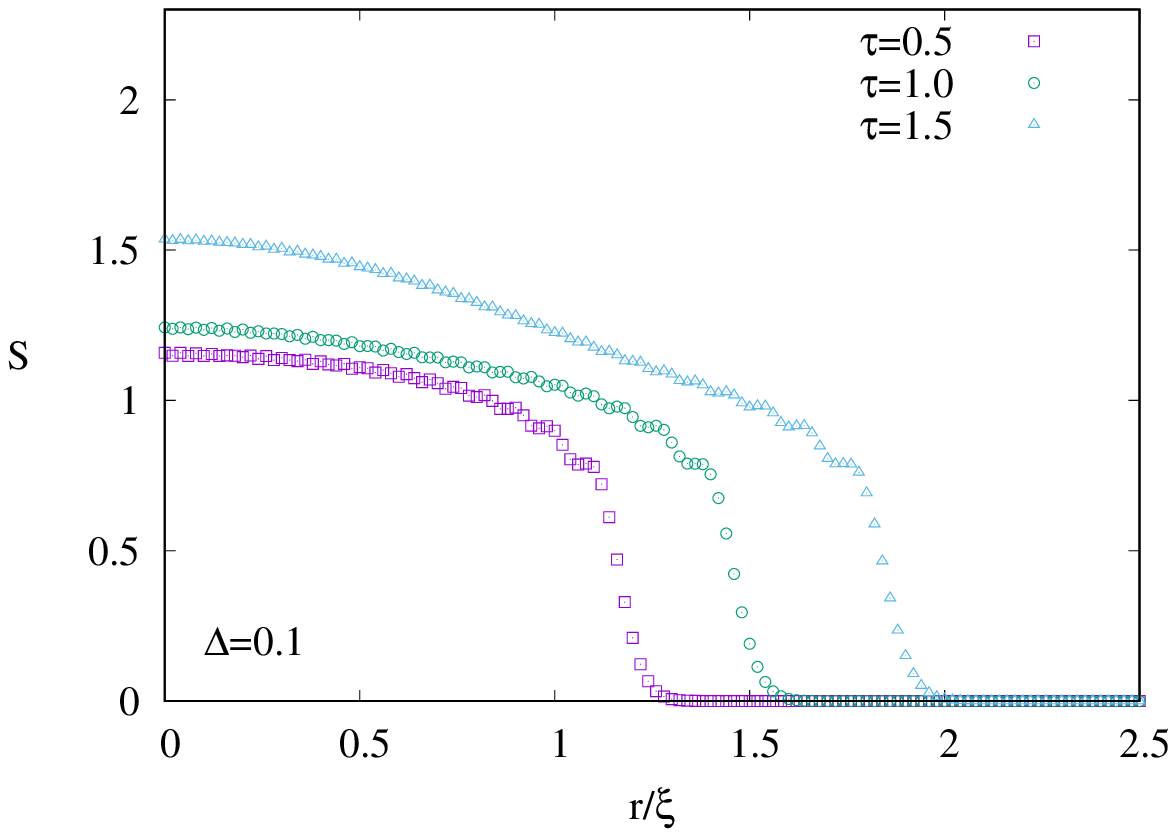}
\includegraphics[width=.66\columnwidth]{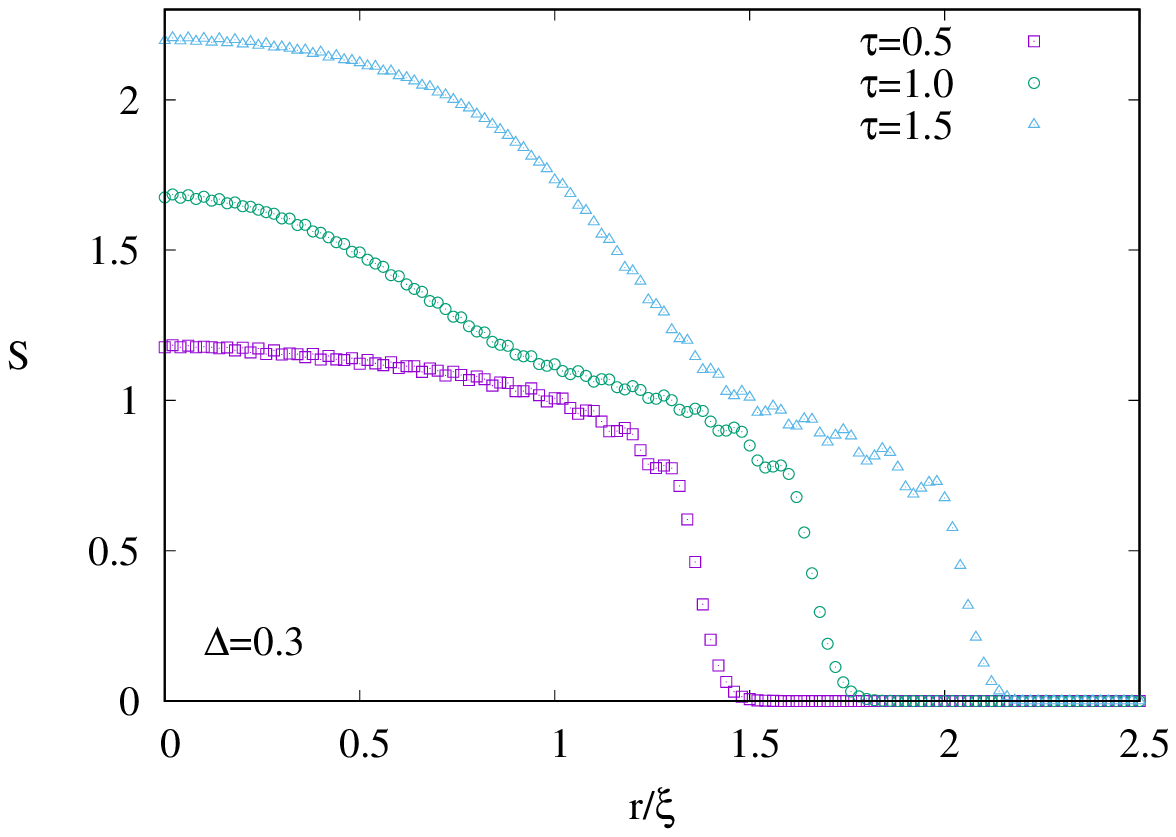}
\includegraphics[width=.66\columnwidth]{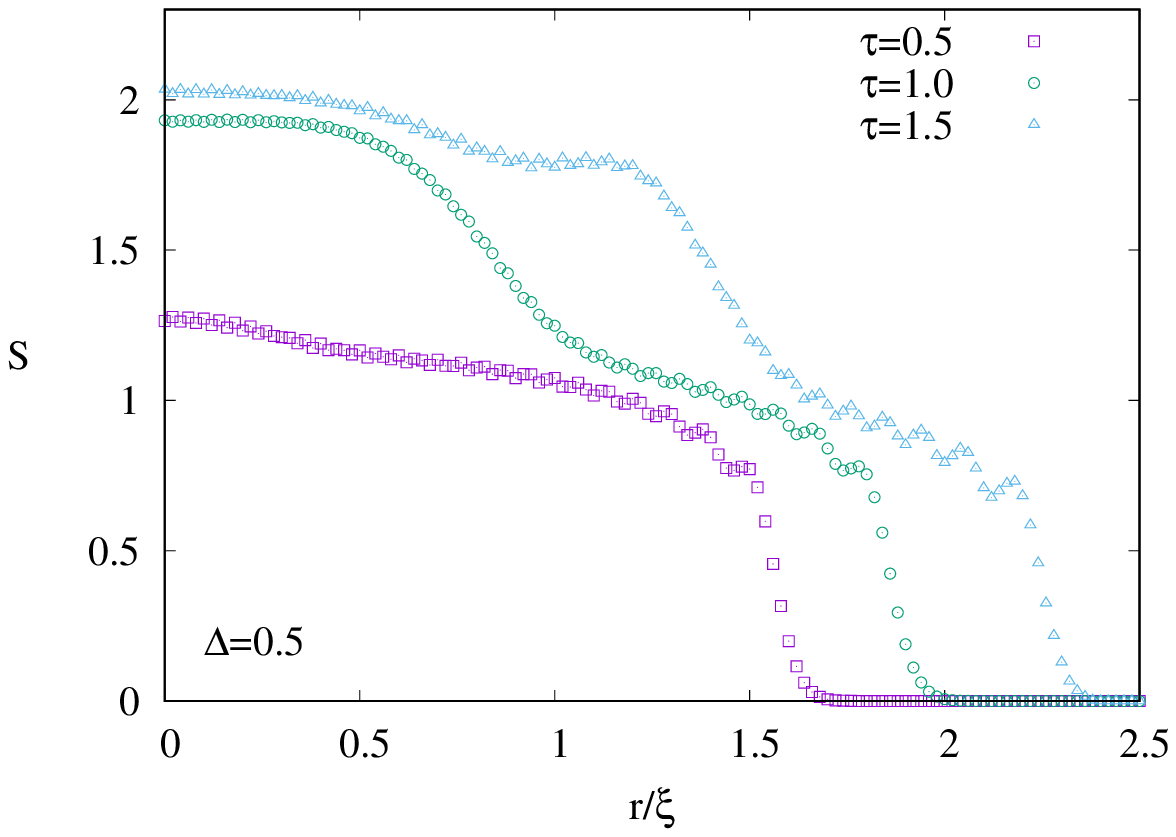}
\caption{Top: magnetization profiles for various times $\tau$, plotted against the
scaling variable $x=j/\xi$. The chain length was set to $L=400$, with $\xi=50$ and
each figure corresponds to a different $\Delta$.
Bottom: entanglement profiles for the same set of parameters,
against the scaled distance $r/\xi$ of the cut measured from the centre.}
\label{fig:szt}
\end{figure*}
%

With the solutions $\Lambda_{\pm}(x,\tau)$ at hand, it remains to extract the
magnetization of the hydrodynamic state with occupation function \eqref{nl}.
Analogously to the ground-state in Eq. \eqref{sz}, it is now given by
\eq{
\langle S^z_j(\tau) \rangle = \frac{1}{2} - \int_{\Lambda_-}^{\Lambda_+}
\tilde\rho (\lambda) \dd \lambda \, ,
\label{szt}}
where $x=j/\xi$ and the root density $\tilde\rho(\lambda)$ follows
from an integral equation similar to Eq. \eqref{rho}, but with the
integration domain replaced by $\left[\Lambda_-,\Lambda_+ \right]$.
The results of this procedure are compared to DMRG calculations
in Fig. \ref{fig:szt} for various $\Delta$. For the time evolution we used
a standard second-order Trotter approximation with time step $\delta t=0.01$
and the discarded weight was set to $10^{-8}$, again without restriction on the
maximal bond dimension. As expected, for small enough
times $\tau$ the hydrodynamic magnetization profile (red lines) show  a very
good agreement with the DMRG data. For larger times, however,
there are visible deviations which tend to increase for larger $\Delta$.

The deviations from the approximate solutions might have different
sources. First of all, we have completely neglected the contributions
from the quasi-particles $l\ne1$ emanating from the left half of the chain.
Indeed, since the evolution of their occupation functions 
is also advective, the spatial domain where $n_l(x,\tau;\lambda)\ne0$
grows for large $\tau$ and eventually penetrates even into the edge region.
For small enough $\tau$, however, due to the boundedness of the
corresponding speed of advection $v_l$, these particles should not contribute
to the magnetization around the edge. Nonetheless, in case of an overlap
region of nonvanishing occupations $l\ne1$ around $x_{min}$, they do couple
the advection equations \eqref{advlpm} via the quasi-particle velocity.
The coupling introduces small changes into the solution $\Lambda_{\pm}(x,t)$
far away from $x_{max}$, which could nevertheless propagate into the
edge region for large enough times. In fact, as remarked in \cite{BVKM17a,BVKM17b},
the proper way to integrate \eqref{adv} would be to apply the method of
characteristics in small backwards time steps
\eq{
n_l(x,\tau;\lambda) = n_l(x-\dd\tau v_l(\{n\};\lambda),\tau-\dd\tau;\lambda) \, ,
\label{bem}}
always
updating the velocities $v_l(\{n\};\lambda)$ using the proper TBA prescription
\cite{BVKM17a,BVKM17b} that takes into account all the occupations $\{n\}$,
evaluated at $x$ and $\tau$.

The effects discussed above can also be visualized by looking at
the entropy profiles, shown in the bottom row of Fig. \ref{fig:szt}.
In fact, there is a clear separation between the edge region,
where the profile resembles that of the ground state (see Fig. \ref{fig:ent}),
and a bulk part with an enhanced amount of entropy. The latter should correspond
to the domain where the quasi-particle excitations from the left
half have already arrived. As time increases, the
edge region gradually shrinks and the separation from the bulk
gets more pronounced for larger $\Delta$. Moreover, for large
$\tau$ and $\Delta=0.5$ one has additional nonmonotonous structures in the profile.

We conclude by providing an approximate formula for
the front size $x_{max}$ in the small time limit $\tau \ll 1$.
As remarked earlier, this is determined by the condition 
$\Lambda_-=\Lambda_+=\Lambda_*$, such that the magnetization
from \eqref{szt} equals $1/2$. Since the occupation is identically
zero, one has no dressing and the velocity is given by
\eq{
v_{*} = 
\frac{\epsilon'(\Lambda_*)}{k' (\Lambda_*)}
\approx \frac{\Lambda_*}{\tan \gamma/2}
}
where in the second step we used that $\Lambda_* \ll 1$
for small times. Using the approximate form of the ground-state magnetization
curve in \eqref{hhc}, the equation \eqref{xlpm} becomes quadratic and 
can be solved for $\Lambda_*$. Requiring the solution to be
unique yields the relation
\eq{
x_{max} \approx h_c + \tau^2/2 \, ,
}
whereas the leading $x \to x_{max}$ behaviour of the magnetization is given by
\eq{
\langle S^z \rangle \approx \frac{1}{2} -  \frac{1}{\pi}\sqrt{2(x_{max}-x)}\, .
\label{szxc}}
%
%
%
The front thus expands very slowly for early times, in complete
analogy with the XX results is \eqref{rhojxt}.
In contrast, for times $\tau \gg 1$ much larger than the inherent length scale
of the problem, one expects the usual ballistic behaviour
$x_{max} \propto \tau$.

\section{Discussion\label{sec:disc}}

We have studied the magnetization and entanglement profiles
for the inhomogeneous ground state of the XXZ chain with a field gradient.
For the description of the magnetization profiles we combined the LDA argument
with the TBA solution in a homogeneous field, and found perfect
agreement for small gradients.
The entanglement profiles are well approximated,
for a wide range of interactions $\Delta$, by a technique
relying on a mapping to a Dirac field theory in curved space.
The dynamics of the interface was also investigated after switching off
the field gradient, and compared to the predictions of generalized
hydrodynamics. For the XX chain the GHD equations can be
solved analytically and reproduce the exact lattice solution in
a proper scaling regime. In the XXZ case we have considered
only an approximate form of the GHD equations which was
argued to be applicable in the edge regime of the front.
The numerical solution of these equations were indeed found
to give a good description of the edge profile for small enough times.

It is very interesting to compare our results to a recent study
of front propagation in the XXZ chain from a domain-wall initial state \cite{CDLV17}.
Due to the simple structure and relations between the quasi-particle occupations in the left-
and right-hand side, an analytical solution to the GHD equations, and hence
the corresponding magnetization and current profiles could be found \cite{CDLV17}.
In particular, it has been shown that, for any $\Delta\ne0$, the magnetization profile around
the front edge changes from a square-root to a linear behaviour.
This is in contrast with our results for the front dynamics from an initial state
with a smooth gradient, where the edge of the profiles still show a characteristic
square-root singularity. For very short times $\tau \ll1$, this behaviour is
also supported by the approximate solution \eqref{szxc} via the GHD equations.
Note, however, that for larger times there is a strong indication of a shrinking
edge regime (see Fig. \ref{fig:szt}) from the DMRG results. Thus it seems
plausible, that for very large times the above regime is entirely washed
out and replaced by a different edge behaviour. To check this one should
push the simulations further in time, which requires considerable efforts.

It would also be desirable to obtain a complete numerical solution of the 
GHD equations \eqref{adv}, instead of the ansatz \eqref{nl} which is only
applicable around the front edge for small enough times.
For an initial state with a slowly varying temperature profile \cite{BVKM17a,BVKM17b},
the integration of the advection equations was achieved by a backwards Euler method, Eq. \eqref{bem},
which is essentially the method of characteristics applied repeatedly with an infinitesimal time step.
This could be generalized to the present case by taking into account the proper
initial conditions $n_l(x,0;\lambda)$ for the full set of occupation functions.
The question is, however, if the occupation functions characterized by a single Fermi sea
would be stable under such an advective evolution. In fact, in a recent study of the Lieb-Liniger
model with an inhomogeneous initial density profile \cite{DDKY17}, it has been pointed out
that instabilities are likely to occur, leading to a breakup of the Fermi sea into several
disconnected parts. 
Whether this would also occur for the gradient XXZ chain at
hand is clearly a very interesting open question which deserves further studies.

Another open issue is the change in the transport properties
as one approaches $\Delta \to 1$, i.e. the boundary of the gapless regime.
Indeed, for the evolution from a sharp domain wall, it was
observed in earlier numerical studies that the ballistic behaviour
is replaced by a slower, but still super-diffusive transport \cite{GKSS05}.
Although the super-diffusivity seemed to be confirmed by further numerical studies \cite{LZP17a,LZP17b},
very recently it has been argued that the dynamics of the domain-wall melting
could also be compatible with simple diffusion \cite{Stephan17,MMK17}.
Studying the melting of a smooth interface in the gradient XXX chain
could shed further light to this question.

Regarding the entanglement profiles, it should be stressed that our curved-space CFT
arguments for the calculation of the ground-state entropy apply only in a perturbative
regime. However, we identified that the relevant small parameter $D(x) \ll 1$ is the scaling
dimension in Eq. \eqref{D}, whereas the Luttinger parameter itself might still show
considerable deviations from the free-fermion point $K=1$.
Nevertheless, it would be of great interest to devise a method
which could incorporate an arbitrary smooth variation of the Luttinger parameter $K(x)$
(i.e. the compactification radius of the CFT). Note, however, that even if this problem
could be tackled, a complete analytical solution for the entropy profile seems out of
reach for $\Delta\ne0$, as it would require knowledge of the non-universal constant term 
in \eqref{shom} for a homogeneous XXZ chain. While this term is easy to extract from
data fits (see Fig. \ref{fig:c1} in Appendix \ref{app:cft}), we are not aware of any
analytical approach to this problem.


Finally, understanding the dynamical entropy profiles also remains an open question.
Our numerical results make it clear that there is strong qualitative difference
between the XX and XXZ chains in this respect. Indeed, in the non-interacting case
one deals essentially with a ground-state problem, with the full entropy profile
given as in Eq. \eqref{entxx} under substitution of an effective length scale
$\sqrt{\xi^2+t^2}$. In contrast, for $\Delta\ne0$
there is a clear separation between the bulk and edge regimes. While the latter still
has the qualitative features of a ground-state profile, the bulk shows a more rapid
growth of entanglement, with the separation becoming more pronounced for larger $\Delta$.
In view of the recent analytic results on the entropy evolution in XXZ chains \cite{AC17,Alba17},
the question naturally emerges whether similar arguments could be applied 
for the present case.

\begin{acknowledgments}

We thank P. Calabrese, M. Kormos and B. Pozsgay for useful discussions.
The authors acknowledge funding from the Austrian Science Fund (FWF) through
projects No. M1854-N36 and P30616-N36, and through SFB ViCoM F41, project P04.

\end{acknowledgments}

\appendix
\section{$v_F$ and $K$ from TBA\label{app:vfk}}

In this appendix, we present numerical results on the
magnetic field-dependence of the Fermi velocity $v_F(h)$
and Luttinger parameter $K(h)$ in the ground state of the
XXZ chain. The calculations are based on the TBA formalism
and follow the lines of Refs. \cite{MT,FF}.

For the sake of completeness, we start with the derivation
of the formula \eqref{hlam}, relating the magnetic field
to the rapidity parameter $\Lambda$. The value of $h$ is set by
the condition $\epsilon^{dr}(\Lambda)=0$, i.e. the dressed
energy of the ground state must vanish at the Fermi rapidity.
The dressed energy follows from the integral equation
\eq{
\epsilon^{dr}(\lambda) + \frac{1}{2\pi} \int_{-\Lambda}^{\Lambda} 
\mathcal{K}(\lambda-\mu) \epsilon^{dr}(\mu) \dd \mu = \epsilon(\lambda) \, ,
\label{edr}}
where the bare energy is given by
\eq{
\epsilon(\lambda) = h -
\frac{\sin^2 \gamma}{\cosh \lambda -\cos \gamma} \, .
\label{epsl}}
Using the TBA equations for the root density \eqref{rho}
and dressed charge \eqref{Z}, the dressed energy can be
rewritten as
\eq{
\epsilon^{dr}(\lambda) = h Z(\lambda) - 
2\pi\rho(\lambda) \sin \gamma \, .
}
Setting $\epsilon^{dr}(\Lambda)=0$, one immediately obtains
\eqref{hlam} of the main text.

The Fermi velocity is defined as the derivative of the dressed energy
w.r.t. the dressed momentum
\eq{
v_F = \left.\frac{\partial \epsilon^{dr}(\lambda)}
{\partial k^{dr}(\lambda)} \right|_{\Lambda} ,
}
evaluated at the Fermi rapidity. Introducing
$v^{dr} = (\epsilon')^{dr}$ with the prime denoting derivative w.r.t. rapidity,
the dressed velocity follows from the integral equation
\eq{
v^{dr}(\lambda) + \frac{1}{2\pi} \int_{-\Lambda}^{\Lambda} 
\mathcal{K}(\lambda-\mu) v^{dr}(\mu) \dd \mu = \epsilon'(\lambda) \, .
\label{vdr}}
Moreover, noticing that the derivative of the bare momentum is given by
$k'(\lambda)=\theta'_1(\lambda)$ as defined in Eq. \eqref{tpn}, 
it follows from \eqref{rho} that the derivative of the dressed momentum
is simply the root density multiplied by a factor of $2\pi$. Hence, the
Fermi velocity can be obtained as
\eq{
v_F = \frac{v^{dr}(\Lambda)}{2\pi \rho(\Lambda)} \, .
\label{vf}}

Finally, one should also fix the Luttinger parameter.
The easiest way is to consider the magnetic susceptibility which
can be given as \cite{Sirker12}
\eq{
\frac{\partial \langle S^z \rangle}{\partial h} = \frac{K}{\pi v_F} \, .
}
Together with \eqref{sz} and \eqref{vf}, this could be used
as the definition of $K$. However, using some variational properties
of the Bethe ansatz root density \cite{FF},
it is also possible to show that the Luttinger parameter is simply related to
the dressed charge as
\eq{
K = Z^2(\Lambda) \, .
\label{K}}

The formulas \eqref{vf} and \eqref{K} can now be evaluated, for a fixed
value of $\Lambda$, via the numerical solution of integral equations
\eqref{rho}, \eqref{Z} and \eqref{vdr}. Varying the value of $\Lambda$,
one obtains the Fermi velocity $v_F(\Lambda)$ and Luttinger parameter
$K(\Lambda)$ as functions of the Fermi rapidity.
The functions $v_F(h)$ and $K(h)$ can then be found
by inverting the function $h(\Lambda)$ in \eqref{hlam}.
The result is shown in Fig. \ref{fig:vfk} for various values of the interaction
parameter $\Delta$. The Fermi velocity (shown on the left) is found to be
very well approximated by the square root expression \eqref{vfh} reported
in the main text, although deviations increase with increasing $|\Delta|$.


\section{R\'enyi entropies and curved-space CFT\label{app:cft}}

Here we present the derivation of the entanglement
entropy in the curved-space CFT framework, following closely
the steps of Ref. \cite{DSVC17}.
The standard procedure is to make use of the replica trick and
calculate the Renyi entropies
\eq{
S_n = \frac{1}{1-n} \ln \Tr \, \rho^{n}_A \, .
\label{sn}}
The von Neumann entropy, discussed in the main text, is then
obtained by taking the limit $S=\lim_{n\to 1}S_n$.

For a general CFT, the trace appearing in \eqref{sn} can
be represented as a path integral on an $n$-sheeted Riemann surface.
Instead of calculating partition functions on this complicated manifold,
one can introduce replicated fields and calculate the expectation values
on the complex plane instead \cite{CC09}. The $n$ copies of $\rho_A$ then
have to be correctly sewn together along the edges of the cut on the complex plane,
representing the subsystem $A$. This can be achieved by applying
the so-called twist fields $\mathcal T_n$ (resp. $\bar{\mathcal{T}}_n$), that (anti)cyclically
permute the replicated fields. Expectation values can then be rewritten by appropriate
insertions of twist fields at the locations of the subsystem boundaries.

In particular, for the simple bipartition considered in the text, with only
one contact point at spatial coordinate $r$ between the two halves of
the system, the R\'enyi entropy can be calculated as the expectation
value of a single twist field
\eq{
S_{n} = \frac{1}{1-n} \ln \, \twn{w_0}_{\mathrm{curved}} \, .
}
The subscript shows that the expectation value is to be evaluated
for a CFT defined in a curved space, as discussed in section \ref{sec:cft}.
The insertion point $w_0 =w(r,0)$ corresponds to the transformed
complex coordinate \eqref{wxy} of the boundary at zero imaginary time.
To evaluate expectation values in the curved geometry, let us
observe that the deformed metric in \eqref{ds2} corresponds to a
simple Weyl transformation, i.e. a stretching of the coordinates
in the $x$-direction. The expectation values can then be related
to those in a flat metric as
\eq{
\twn{w_0}_{\mathrm{curved}} =
\ee^{-\sigma\Delta_n} \twn{w_0}_{\mathrm{flat}} \, ,
}
where the scaling dimension of the twist field is \cite{CC09}
\eq{
\Delta_n = \frac{c}{12}(n-1/n) \, ,
}
and $c$ is the central charge of the CFT.

The last step is to evaluate the twist-field one-point function
in the flat metric. Recall that the CFT in the original $z$ complex coordinates
lives on an infinite strip, with $\rp z = x \in \left[-R,R\right]$ where
$R=h_c \xi$. In the stretched $w$ coordinates the strip domain is
given by $\rp w \in \left[-W,W\right]$ where
\eq{
W = w(R,0) = \int_{0}^{R} \frac{\dd x'}{v_F(x')} \, .
}
The infinite strip can be mapped to the upper half plane by
the simple conformal mapping
\eq{
w \to g(w) = \ee^{i\pi\frac{w+W}{2W}} \, .
\label{gw}}
Under this conformal map, the expectation values transform as
\eq{
\twn{w_0}_{\mathrm{flat}} = 
\left|\frac{\dd g}{\dd w}\right|^{\Delta_n}_{w_0} \twn{g(w_0)}_{\mathrm{uhp}} \, .
}
%
%
\begin{figure}[htb]
\center
\includegraphics[width=\columnwidth]{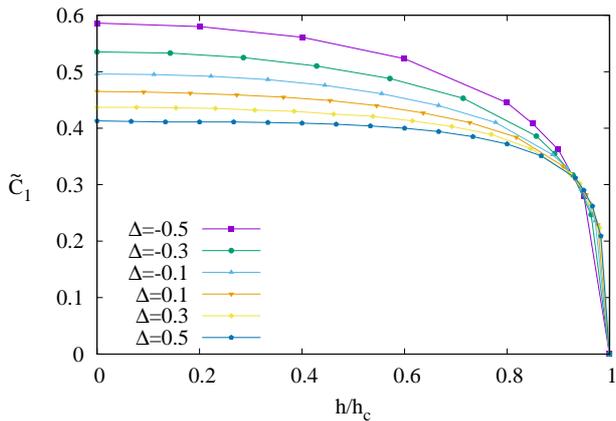}
\caption{Non-universal term in the ground-state entanglement entropy, as a result of
fitting the ansatz \eqref{sngs} with $n=1$ and $c=1$. The length of the chain was $L=200$.}
\label{fig:c1}
\end{figure}
%
%
Finally, the expectation value on the upper half plane is simply given by
\eq{
\twn{g(w_0)}_{\mathrm{uhp}} \propto (\ip g(w_0))^{-\Delta_n} \, ,
}
where the proportionality sign indicates the presence of
non-universal multiplicative factors.

Putting everything together, we arrive at the following expression for the
R\'enyi entropy
\eq{
S_n = \frac{c}{12}(1+1/n) \ln \mathcal L + C_n \, ,
\label{sn2}}
where the conformal distance is given by
\eq{
\mathcal{L} = \ee^{\sigma} \left|\frac{\dd g}{\dd w}\right|^{-1}_{w_0}\ip g(w_0) \, .
\label{l1}}
Substituting $\ee^{\sigma}=v_F$ and using \eqref{gw}, this can
be rewritten as
\eq{
\mathcal{L} = v_F \frac{2W}{\pi} \cos
\left(\frac{\pi}{2W} \rp w_0\right) .
\label{l2}}
Now the term $\rp w_0$ is given by an integral expression in terms of
the Fermi velocity, for which we do not have an exact analytical
expression. However, we have found that the simple square-root
formula in Eq. \eqref{vfh} gives a very good approximation of the
actual data obtained from TBA. Using this approximation, the integral
can be carried out and one finds
\eq{
\rp w_0 = \frac{R}{v_0} \arcsin \left( \frac{r}{R} \right) .
}
Inserting this into \eqref{l2}, we obtain the conformal distance as
\eq{
\mathcal{L} = R\left[1-\left(\frac{r}{R}\right)^2 \right] .
\label{l3}}
Using $R=h_c \xi$, we arrive at the formula \eqref{l} reported in the main text.

Unfortunately, due to the presence of the non-universal term $C_n$,
the expression in \eqref{sn2} does not fix the full analytical
form of the R\'enyi entropies. In fact, this term still depends on both
the location of the cut $r$, as well as the interaction parameter $\Delta$.
It is, however, possible to fix $C_n$ from ground-state
calculations of the entropy on the lattice.
Indeed, considering a finite XXZ chain of length $L$ in a \emph{homogeneous}
magnetic field $h<h_c$, the ground state corresponds to a $c=1$ CFT
on a strip. Using the mapping \eqref{gw} with the substitution
$W \to L/2$, the CFT result for the entropy of a segment of length $\ell$ reads
\eq{
S_n = \frac{c}{12}(1+1/n)  \ln \tilde{\mathcal{L}} + \tilde C_n \, ,
\label{sngs}}
where the conformal distance is given by the chord length in Eq. \eqref{lgs}.
Hence, setting the ground-state magnetic field equal to $h=r/\xi$,
one finds from analogy of Eqs. \eqref{sn2} and \eqref{sngs}
\eq{
C_n(r,\Delta) = \tilde C_n(r/\xi,\Delta) \, .
}

The variation of $C_n$ as a function of the cut position can thus be inferred
from the ground-state behaviour $\tilde C_n$ as a function of the magnetic field.
The latter can be obtained by analyzing the lattice data obtained from DMRG
for various $\ell \le L/2$, with fixed $h,\Delta$ and $L$,
and fitting the ansatz \eqref{sngs} for each data set. Repeating the procedure
for various $h$ and $\Delta$, we obtained the non-universal term for the $n=1$ case.
This is shown on Fig. \ref{fig:c1}, as a function of the scaled magnetic field
$h/h_c$ and various values of $\Delta$. It should be noted that an analytical
expression of $\tilde C_1$ is known only for the free-fermion case $\Delta=0$,
where it is given by Eq. \eqref{cxx}. Numerical results for the case $h=0$
were also obtained in Ref. \cite{CCEN10}.

\bibliographystyle{apsrev.bst}

\bibliography{xxzgrad_refs}

\end{document}